\documentclass[aps,pre,twocolumn,showpacs,superscriptaddress]{revtex4-1}

\usepackage{amssymb,amsmath}
\usepackage{hyperref} 
\usepackage{graphicx}

\usepackage{dcolumn} 
\newcommand{\specialcell}[2][c]{\begin{tabular}[#1]{@{}c@{}}#2\end{tabular}} 

\usepackage{siunitx} 
\usepackage{physics} 
  \renewcommand{\v}[1]{\vb*{#1}} 

\newcommand{\inv}{^{-1}} 
\newcommand{\e}[1]{\, \mathrm{e}^{#1}} 
\newcommand{\nn}[1]{\langle #1 \rangle} 

\DeclareMathOperator*{\Var}{Var}

\usepackage{mathtools}
\usepackage[capitalise]{cleveref}

\begin{document}
\title{Policy Guided Monte Carlo: Reinforcement Learning Markov Chain Dynamics}

\author{Troels Arnfred Bojesen}
\email[]{troels.bojesen@aion.t.u-tokyo.ac.jp}
\affiliation{Department of Applied Physics, University of Tokyo, Hongo 7-3-1, Bunkyo-ku, Tokyo 113-8656, Japan}

\date{\today}

\begin{abstract}
  We introduce \textit{Policy Guided Monte Carlo} (PGMC), a computational framework using reinforcement learning to improve Markov chain Monte Carlo (MCMC) sampling. The methodology is generally applicable, unbiased and opens up a new path to automated discovery of efficient MCMC samplers. After developing a general theory, we demonstrate some of PGMC's prospects on an Ising model on the kagome lattice, including when the model is in its computationally challenging kagome spin ice regime. Here, we show that PGMC is able to automatically machine learn efficient MCMC updates without a priori knowledge of the physics at hand.
\end{abstract}

\pacs{02.70.Tt, 02.70.Rr, 02.70.Uu, 05.10.Ln}

\maketitle

\section{Introduction}\label{sec:introduction}
Since the invention of Markov chain Monte Carlo (MCMC) in the mid 20th century~\cite{Metropolis1953,RobertCasella2011}, MCMC methods have grown to not only become a backbone of computational physics, but arguably of modern civilization as a whole~\cite{SullivanDongarra2000}. The basic idea of MCMC is seemingly innocuous: use a Markov chain to obtain a set of samples $\{s\}$ from a state space $\mathcal{S}$, each sample with a probability weight (or density) given by some function, or \emph{model}, $w: \mathcal{S} \to \mathbb{R}_{\ge 0}$. Unfortunately, however, the autocorrelation of the samples means that long and possibly computationally expensive simulations may be needed in order to achieve reliable results. This reality poses a threat to the feasibility of using MCMC in practice, as real world computational resources are limited (and comes with a monetary cost). 

To combat this problem, a plethora of MCMC algorithms, methods, and paradigms have been developed, most of which attempt to improve the original Metropolis-Hastings framework~\cite{Metropolis1953,Hastings1970} in one way or another~\cite{[{See e.g. }] [{ for a non-exhaustive selection of examples.}]LandauBinder2009}. Among those, an intriguing group of algorithms has emerged with the recent influx of machine learning ideas and techniques to the domain of MCMC: The \emph{Effective model Monte Carlo} (EMMC) method~\cite{Note102,PhysRevB.95.035105,PhysRevB.95.041101}\footnotetext[102]{We use the label \emph{Effective model Monte Carlo} instead of e.g. \emph{Self-learning Monte Carlo} as done in \cite{PhysRevB.95.041101}, since the latter is I) a too broad description and II) strongly resembles the name of the algorithmically different (and by-more-than-a-decade preceding) \textit{Self-learning kinetic Monte Carlo} method~\cite{PhysRevB.72.115401}}.

In EMMC, the task of sampling states following model $w$ is divided into two stages: A \textit{learning} or \textit{training} stage, and an \textit{earning} or sampling stage. First, one obtains a set of training data, meaning a set of samples following the distribution of $w$, e.g. by means of a traditional MCMC simulation. Then, an \textit{effective model} $\tilde w_{\v \theta}:\tilde{\mathcal{S}} \to \mathbb{R}_{\ge 0}, \tilde{\mathcal{S}} \supseteq \mathcal{S}$, parametrized by a set of parameters $\v \theta = \{\theta_i\}$, is fitted to this dataset by some form of regression, i.e. \textit{machine learned}, such that $\tilde w_{\v \theta}(s) \approx w(s)$ for states of statistical significance. The effective model should be constructed in a way that makes it computationally advantageous over the original model $w$. It could, for example, be computationally cheaper to evaluate and/or open up for the use of a previously inaccessible sampling method. In the second stage, $\tilde w_{\v \theta}$ is used as a \textit{proposal generator}: A number of updates, $s \to s_1 \to s_2 \to \ldots s_{n-1} \to s_n = s'$, is performed following some MCMC sampling scheme applied to the effective model, $\tilde w_{\v \theta}$. Then, the new state $s'$ is treated as a proposal for the sampling of the original model, in which case it is accepted with Metropolis acceptance rate~\footnote{Or any other acceptance rule fulfilling the detailed balance condition.}
\begin{equation}
 \alpha(s \to s') = \min\left[1, \frac{\tilde w_{\v \theta}(s)w(s')}{\tilde w_{\v \theta}(s') w(s)}\right].
 \label{eq:surrogate_acceptance}
\end{equation}
This ensures that the detailed balance condition
\begin{equation}
w(s)\pi(s \to s')\alpha(s \to s') = w(s')\pi(s' \to s)\alpha(s' \to s)
 \label{eq:detailed_balance}
\end{equation}
holds, in which case the Markov chain will asymptotically sample the distribution of $w$ instead of the distribution of $\tilde w_{\v \theta}$. Here $\pi(s \to s')$ denotes the probability of proposing the change $s \to s'$. By choosing a sensible effective model and update rule, an overall faster diffusion in state space (in CPU and/or real time) can be achieved, with obvious beneficial implications.

Although the EMMC scheme has demonstrated significant speedups in various instances~ \cite{PhysRevB.95.035105,PhysRevB.95.041101,PhysRevE.95.031301,PhysRevB.95.241104,PhysRevB.96.041119,PhysRevB.96.161102,PhysRevB.98.041102,PhysRevB.98.045116}, it is not without caveats and shortcomings. First, the effective model has to be flexible enough to be able to ``imitate'' the original model. If this is not the case, EMMC may perform worse than a traditional, ``naive'' MCMC method. Second, the effective model has to be trained using sufficiently good training data, with complex effective models with more parameters requiring more data. Acquiring the training data can in itself be challenging, e.g. if the dynamics of the MCMC sampler used is slow. Should the training data not represent a sufficiently good sampling of the relevant parts of state space, the effective model, and hence the final outcome, will reflect this. In the best case, this only makes the EMMC simulation slower, in the worst, the proposed states will be drawn from a wrong or incomplete subset of state space. 

Issues concerning the first stage of the EMMC can be somewhat mitigated by iteratively (and gradually) improving the effective model, constructing a cascade of learning runs based on the previously obtained effective models, until convergence has been reached. Such a scheme does, however, not resolve situations where the MCMC algorithm used to generate proposal states based on the effective model is intrinsically inefficient or inappropriate for the problem at hand. For example, an effective model based proposal generator that is \textit{frozen}, in the sense that it never or almost never proposes significant movements in state space, will lead to poor MCMC performance, (almost) regardless of each iteration's evaluation speed. EMMC is only as good as its weakest link: the proposal generation.

Naturally, the latter concerns can sometimes -- when the details of the Markov chain dynamics is known a priori -- be countered by hand crafted updates during the sampling run. Such ``human intervention'' does, however, offset some of the usefulness and automatic discovery potential of having machine learned algorithms in the first place. 

In this work, we take a different approach in the quest for not only finding efficient MCMC algorithms, but also doing so in an as-automatic-as-possible fashion. Instead of focusing on constructing effective models, we argue that it is more fruitful to target the proposal generation directly by tuning the \textit{policy} for making updates, akin to what is done within the \textit{reinforcement learning} branch of machine learning~\cite{SuttonBarto2018}. After all, we are primarily interested in the (efficient) diffusion of Markov chains through state space, not the effective models per se~\footnote{Even if the goal is to determine an effective model, we would claim that PGMC sampling is superior to EMMC in obtaining the necessary training data}.

The idea of automatically tuning the proposal distributions of MCMC samplers is not new: It is not unreasonable to speculate that ad hoc solutions may have emerged already shortly after the invention of MCMC~\footnote{After all, Metropolis et al. themselves note, with respect to their tunable parameter $\alpha$, ``It may be mentioned in this connection that the maximum displacement $\alpha$ must be chosen with some care; if too large, most moves will be forbidden, and if too small, the configuration will not change enough.''~\cite{Metropolis1953} The motivation for finding a way to more-or-less automatically choose a ``decent'' $\alpha$ was present right from the start.}. Later attempts at formalizing the process can be found e.g. in Ref.~\cite{Roberts:1997} (and references therein) as well as in form of the so-called \textit{adaptive MCMC} algorithms~\cite{Note103,Note104}\footnotetext[103]{See for instance Refs.~\onlinecite{RobertsRosenthal2009, Andrieu2008} and references therein. Examples of algorithms showing some similarities with the algorithms presented in this work can be found in e.g. Refs.~\onlinecite{łatuszyński2013} and \onlinecite{Yang2017}.}\footnotetext[104]{The naming is rather confusing, as the adaptive MCMC algorithms are \emph{not} markovian}. Shared for these is the tuning of some defining parameters of select -- typically gaussian -- proposal distributions. In joining forces with reinforcement learning, however, we will show that it is possible to take the idea even further.

In the following section, we will connect a few core concepts of reinforcement learning to MCMC as we prepare the playing ground, before we in \cref{sec:PGMC} develop a general reinforcement learning MCMC framework, dubbed \textit{Policy Guided Monte Carlo} (PGMC). In \cref{sec:PGMC_in_practice}, PGMC is demonstrated through a series of simulations of the kagome lattice Ising model in a field. The examples also open up for a more detailed discussion of the specifics of the methodology. Finally, we attempt to put PGMC somewhat in perspective in \cref{sec:discussion}, before concluding in \cref{sec:conclusion}~\footnote{During the preparation of this manuscript, the author became aware that Ref.~\cite{PhysRevE.96.051301} briefly floats a similar idea of applying reinforcement learning methods to directly search for an optimal policy. The idea is, however, not pursued in any way in Ref.~\cite{PhysRevE.96.051301}.}.

\section{MCMC in light of reinforcement learning}\label{sec:MCMC_in_ligh_of_RL}
The Markov decision processes heavily employed within reinforcement learning may be regarded as a superset of the Markov chains~\cite{SuttonBarto2018}. Thus, a lot of the formalism developed in reinforcement learning can readily be adopted to MCMC: The update $s \to s'$ may be regarded as an \textit{action} -- one of possibly many taking the system, or \textit{agent}, from state $s$ to some other state $s'$. We denote the space of all actions $\mathcal{A}_{s}$, where the subscript $s$, when included, implies that the possible actions may depend on the initial state $s$. The \textit{policy} for taking action $a: s \mapsto s', a \in \mathcal{A}_s$ is quantified by the proposal probability $\pi: \mathcal{A}_{s} \times \mathcal{S} \to [0,1]$. We will later refer to $\pi$ itself as ``the policy''. The \textit{inverse action} is denoted $a\inv : s' \mapsto s$.

In Metropolis-Hastings MCMC, the acceptance probability $\alpha$ can, in principle, be chosen in any way that satisfies Eq.~\eqref{eq:detailed_balance}. However, for simplicity, and in order to make the discussion more concrete, we will be using the Metropolis acceptance rate~\cite{Metropolis1953,Hastings1970}
\begin{equation}
 \alpha(s \to s') = \min{\left[1, \frac{w(s')\pi(s'\to s)}{w(s)\pi(s \to s')}\right]}.
 \label{eq:metropolis_acceptance}
\end{equation}
Then, the Markov chain dynamics will be determined solely by the choice of policy $\pi$ (which implicitly also determines the action space), as the model $w$ is given a priori~\footnote{The derivations done here can straightforwardly be adopted to other acceptance rates as well. In case these acceptance rates involve tunable parameters, like in Ref.~\cite{MuellerKrumbhaarBinder1973}, the parameters can be treated similarly to other parameters in the PGMC algorithm [see \cref{sec:PGMC}].}.

In the original Metropolis algorithm, the policy is symmetric, $\pi(s \to s') = \pi(s' \to s)$, rendering Eq.~\eqref{eq:metropolis_acceptance} particularly simple. But although convenient, there is no reason to believe that this choice is optimal. For example, by associating $\pi(s \to s')$ with updates based on $\tilde w_{\v \theta}$, the proposal generation of EMMC may be seen as a (typically) asymmetrical policy. A similar case can be made for many other existing MCMC algorithms.

\begin{figure}
  \includegraphics[width=\columnwidth]{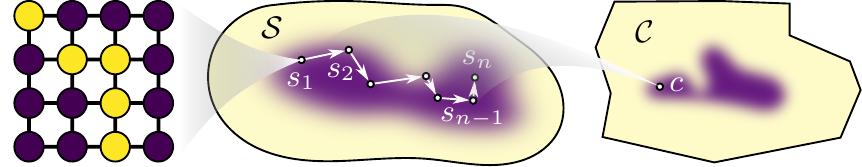}
  \caption{A sketch of the hierarchy of abstractions in play: A state, e.g. a configuration of a lattice model, constitutes a single point in a high dimensional state space $\mathcal{S}$. A trajectory of states $s_1 \to s_2 \to \ldots \to s_n$ of length $n$ corresponds to a single point $c$ in the even higher dimensional trajectory space $\mathcal{C}$. The darker areas illustrate the regions of higher probability density that contributes the most to the final result.}
 \label{fig:states_and_trajectory}
\end{figure}

A natural question then follows: What is the optimal policy for an MCMC simulation of a given model? To address this in a quantitative way, we look at the sampling of some observable $O$. (After all, this is ultimately the goal of the simulation.) The statistical efficiency of this process is typically gauged by the associated integrated autocorrelation time $\tau_{O}: \mathcal{C} \to [\frac{1}{2},\infty)$~\cite{Sokal1997, Note101}.\footnotetext[101]{In this work, we assume that the samples may not be anticorrelated. Nevertheless, even if (mild) anticorrelation emerge, the algorithms and procedures presented are expected to qualitatively work} Here, $\mathcal{C}$ denotes the set of all possible \textit{trajectories} of states the Markov chain can follow through state space, as illustrated in \cref{fig:states_and_trajectory}. In this work, $\tau_O$ is measured in units of Monte Carlo updates, so to reflect the real world performance of generating a trajectory, the autocorrelation time should be multiplied by a \textit{cost factor} $u: \mathcal{C} \to \mathbb{R}_{>0}$ that -- ideally -- incorporates all costs of importance associated with obtaining a trajectory. The ``costs of importance'' could, for example, be the time or computational resources needed. 

Equivalently -- and more conveniently -- we introduce the \textit{performance factor}
\begin{equation}
 \epsilon_O(c) \equiv \frac{1}{2\tau_{O}(c) u(c)} > 0,\quad c\in \mathcal{C}
 \label{eq:pf}
\end{equation}
as a measure of efficiency; the larger the performance factor, the better the MCMC sampling. If $\epsilon_O(c) = 0$, all samples are perfectly correlated and/or infinitely expensive to generate, while $\epsilon_O(c) = u(c)\inv$ means that they are statistically independent. Thus, all other factors being equal, the optimal policy, $\pi^*$, is the one that maximizes the expected performance factor,
\begin{equation}
 \ev{\epsilon_O}_{c\sim w} = \sum_{c \in \mathcal{C}} \epsilon_O(c)p(c).
 \label{eq:pf_expectation}
\end{equation}
The subscript $c \sim w$ (or sometimes just $\sim w$) signifies that the states of the trajectory $c$ are sampled following the $w$ distribution. This statistics is captured by $p(c)$, the probability of creating $c$. In the language of reinforcement learning, the performance factor plays the role of a \textit{reward function}. Denoting by $\pi(c)$ and $\alpha(c)$ the probability of proposing and accepting a trajectory $c = (s_1 \to s_2 \to \ldots \to s_n)$ of some length $n$, we have
\begin{equation}
 \begin{split}
  p(c) &= p(s_1)\pi(c)\alpha(c) \\
  &= p(s_1)\prod_{t=1}^{n-1} \pi(s_t \to s_{t+1})\alpha(s_t \to s_{t+1})
 \end{split}
 \label{eq:trajectory_probability}  
\end{equation}
due to the Markov property of the Markov chain. $p(s_1)$ is the probability of starting in state $s_1$, which must fulfill $p(s_1) \propto w(s_1)$ if $c \sim w$ is to hold in \cref{eq:pf_expectation}. (In practice, this will be the case if the trajectories are obtained from equilibrated MCMC simulations.)

\section{Policy guided Monte Carlo}\label{sec:PGMC}
Similar to EMMC, we propose a two stage PGMC scheme:
\begin{enumerate}
 \item Given a model $w$, find a policy $\pi^*$ that makes \cref{eq:pf_expectation} as large as possible.
 \item Use $\pi^*$ in an MCMC simulation to obtain samples distributed according to probability weight $w$.
\end{enumerate}
Thus, unlike conventional reinforcement learning, where the policy determines the complete behavior of an agent, in PGMC the policy only \textit{guides} the behavior of the Markov chain. The final ``judge'' is always the MCMC acceptance step, which enforces the sampling of $w$ to be unbiased. Importantly, this holds regardless of the details of the policy, as long as it is ergodic. 

Whereas stage 2 is conceptually straightforward, stage 1 demands a more throughout treatment. In the next subsections, we will discuss each of the following central aspects of finding $\pi^*$: policy search, how to model the policy, and how to estimate the performance factor.

\subsection{Policy search}\label{sec:policy_search}

Any practical attempt at finding $\pi^*$ will run into at least two obstacles: I) The set of all possible policies, $\mathcal{P}$, is infinite. II) The set of all possible trajectories, $\mathcal{C}$, is either infinite or extremely large. Clearly, approximations has to be made in order to proceed. We should therefore not expect to find \emph{the} optimal policy, but rather ``just'' a good one.

To tackle I), we restrict the policy search to $\{\pi_{\v \theta}\} \subset \mathcal{P}$ for some \textit{policy model} $\pi_{\v \theta}$ parametrized by $\v \theta \in \mathbb{R}^d$. Demanding $\pi_{\v \theta}$ to be (almost everywhere) differentiable with respect to $\v \theta$, we may proceed by policy gradient optimization~\cite{SuttonBarto2018}. At a conceptual level, the procedure is to perform gradient descent on $-\ev{\epsilon_O}_{c\sim w}$, now a function of $\v \theta$, until a minimum is reached. In practice, we will use some variant of \textit{stochastic gradient descent} to do so~\cite{BottouEtAl2018}.

Using stochastic gradient decent also partially solves II), as the optimization is now based on a stochastic sampling of $\mathcal{C}$, rather than a practically impossible exact integration over $\mathcal{C}$. The samples, i.e. trajectories, can be generated and stored as a set of training data prior to the optimization (\textit{off\-line} optimization), although in reinforcement learning it is typically more advantageous to generated them on-the-fly, as needed (\textit{online} optimization). Here, we will focus on the latter approach.

\begin{figure}
  \includegraphics[width=\columnwidth]{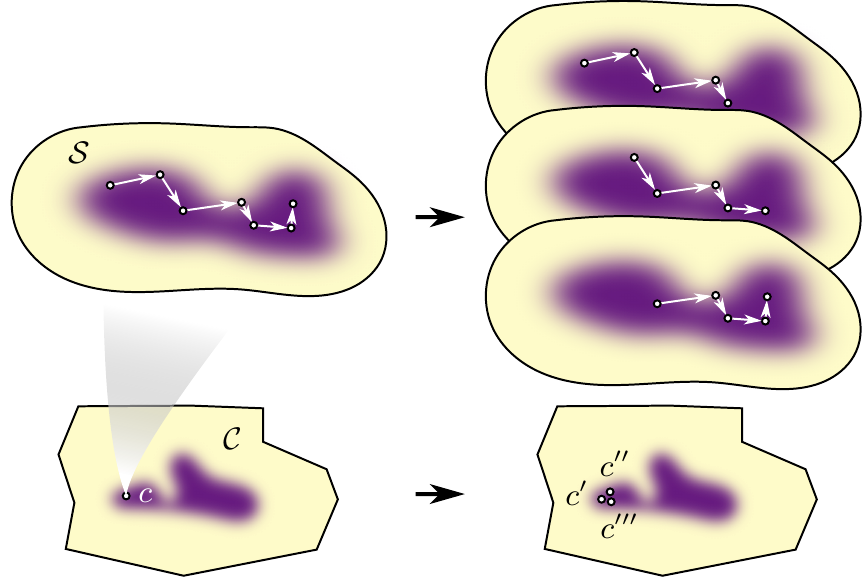}
  \caption{The trajectory $c$ may be regarded as a number of shorter, time-shifted, and correlated subtrajectories, e.g. $c'$, $c''$, and $c'''$.}
 \label{fig:subtrajectories}
\end{figure}

Generating even a single trajectory can be expensive if we demand it to be statistically independent of previous trajectories. Fortunately, this is not necessary for the stochastic gradient decent to converge; as long as all regions of $\mathcal{C}$ of importance are eventually visited, it is acceptable with correlations between subsequent samples. This, together with the statistical time translation invariance of the Markov chain in equilibrium, means that time shifted segments of the same trajectory may be regarded as separate, shorter (and strongly correlated) trajectories, or \textit{subtrajectories}; see \cref{fig:subtrajectories}. In this sense, a new trajectory is created with every update step. Furthermore, if old subtrajectories are discarded from the training data as new ones are generated (as is the case with online optimization), it is reasonable to expect that an stochastic gradient decent optimization based on a small set of memorized subtrajectories will converge to a correct optimum, even if initial subtrajectories are out of equilibrium.

During the policy optimization, the trajectories are generated following a \textit{behavior policy} $b \in \mathcal{P}$. The behavior policy may, in general, be different from the \textit{target policy} $\pi_{\v \theta}$ we are trying to optimize. In other words, $b$ controls the update dynamics during the first stage of the PGMC simulation, while the target policy with optimized parameters, $\v \theta = \v \theta^*$, $\pi_{\v \theta^*} \approx \pi^*$, is to be used in the MCMC sampling in the second stage. Formally, \cref{eq:pf_expectation} in conjunction with \cref{eq:trajectory_probability} may be written
\begin{equation}
 \ev{\epsilon_O}_{c \sim w} = \sum_{\substack{s_1 \in \mathcal{S} \\ c \sim b}} \epsilon_O(c)p(s_1)\frac{\pi_{\v \theta}(c)}{b(c)}\alpha_{\v \theta}(c)
 \label{eq:importance_sampling_epsilon}
\end{equation}
since we are generating trajectories based on $b$, as indicated by $c\sim b$, but ultimately desire the statistics generated by $\pi_{\v \theta}$. Like before, each trajectory starts in some state $s_1$. A subscript $\v \theta$ has been included in $\alpha_{\v \theta}$ as a reminder that the acceptance rate also depends on $\v \theta$ [through $\pi_{\v \theta}$; see Eq.~\eqref{eq:metropolis_acceptance}]. We opt to not include a Metropolis acceptance step in the training stage of the simulation (i.e. all updates are accepted), as we desire as fast an exploration of the state space as possible. Then $\alpha_{\v \theta}$ should be treated just as a numerical factor during the optimization. 

Since the terms of \cref{eq:importance_sampling_epsilon} may span several orders of magnitude, especially in the initial iterations of the optimization, we will attempt at maximizing $\ln \ev{\epsilon_O}_{c \sim w}$ rather than \cref{eq:importance_sampling_epsilon} directly. This does not change $\pi^*$, but reduces the variance of the policy gradient estimate, because
\begin{equation}
 \grad_{\v \theta} \ln \ev{\epsilon_O}_{c \sim w} = \frac{\grad_{\v \theta} \ev{\epsilon_O}_{c \sim w}}{\ev{\epsilon_O}_{c \sim w}}
 \label{eq:gradient_log_pf}
\end{equation}
is numerically more stable when $\ev{\epsilon_O}_{c \sim w}$ is approximated by a finite number of stochastically sampled terms~\cite{SuttonBarto2018}.

Moreover, exploiting that a reversible Markov chain in equilibrium is invariant with respect to time reversal, we may perform the replacement
\begin{equation}
 p(c) \to \sqrt{p(\overset{\to}{c})p(\overset{\gets}{c})}.
 \label{eq:time_reversal_symmetrized}
\end{equation}
The arrows in $\overset{\rightleftarrows}{c} = (s_1 \rightleftarrows s_2 \rightleftarrows \ldots \rightleftarrows s_n)$ indicate whether the trajectory $c$ is traced forward or backward in time. Although not imperative, this symmetrization improves the stability and convergence of the optimization. First, the amount of information taken into account is increased by imposing the time reversal symmetry. Second, the multiplicative combination of $p(\overset{\to}{c})$ and $p(\overset{\gets}{c})$ has a regularizing effect on the optimization, since before equilibrium has been reached, a large $p(\overset{\to}{c})$ will not necessarily be favored if it leads to a small $p(\overset{\gets}{c})$. Thus, we expect the symmetrization to be of particular importance in dampening temporary learning of any transient behavior the system might experience before reaching equilibrium. 

Using \cref{eq:metropolis_acceptance,eq:importance_sampling_epsilon,eq:time_reversal_symmetrized}, the symmetry $\epsilon_O(\overset{\to}{c}) = \epsilon_O(\overset{\gets}{c}) = \epsilon_O(c)$, and the identites $\grad_{\v \theta}g_{\v \theta} = g_{\v \theta} \grad_{\v \theta}\ln g_{\v \theta} $ and $\min(0,x) + \min(0,-x) = -\abs{x}$, we get 
\begin{widetext}
 \begin{gather}
  \begin{split}
  \grad_{\v \theta} \ev{\epsilon_O}_{c \sim w} 
  &= 
  \sum_{\substack{s_1 \in \mathcal{S} \\ c \sim b}} p(s_1)\epsilon_O(c)\frac{\pi_{\v \theta}(c)}{b(c)}\alpha_{\v \theta}(c) \grad_{\v \theta} \left[\ln \pi_{\v \theta}(c) + \ln \alpha_{\v \theta}(c)\right] \\
  &\to 
  \frac{1}{2}\sum_{\substack{s_1 \in \mathcal{S} \\ c \sim b}} \sqrt{p(s_1)p(s_n)} \epsilon_O(c)\sqrt{\frac{\pi_{\v \theta}(\overset{\to}{c})\pi_{\v \theta}(\overset{\gets}{c})}{b(\overset{\to}{c})b(\overset{\gets}{c})}} \e{-\frac{1}{2}\abs{\Delta f_{\v \theta}(c)}}\grad_{\v \theta} \left[\ln \pi_{\v \theta}(\overset{\to}{c}) + \ln \pi_{\v \theta}(\overset{\gets}{c}) - \abs{\Delta f_{\v \theta}(c)}\right],
  \end{split}
  \label{eq:general_objective_gradient}
 \end{gather}
where
\begin{gather}
 \begin{split}
  \Delta f_{\v \theta}(c) &\equiv \sum_{t=1}^{n-1}\left[\ln w(s_{t+1}) + \ln \pi_{\v \theta}(s_{t+1}\to s_{t})\right] - \left[\ln w(s_{t}) + \ln \pi_{\v \theta}(s_{t} \to s_{t+1})\right]\\
 &= \ln w(s_n) - \ln w(s_1) + \sum_{t=1}^{n-1}\left[\ln \pi_{\v \theta}(s_{t+1}\to s_{t}) - \ln \pi_{\v \theta}(s_{t} \to s_{t+1})\right] 
 \end{split}
 \label{eq:delta_f}
\end{gather}
\end{widetext}
and 
\begin{equation}
 \grad_{\v \theta} \ln \pi_{\v \theta}(\overset{\rightleftarrows}{c}) = \sum_{t=1}^{n-1} \grad_{\v \theta} \ln \pi_{\v \theta}(s_t \rightleftarrows s_{t+1}) .
\end{equation}

\Cref{eq:general_objective_gradient} is the most general policy gradient for PGMC policy optimization. It can, in principle, be employed using any ergodic behavior policy $b$ (\textit{off-policy} learning), but unless a good behavior policy can be found, bad or very bad convergence is to be expected~\cite{SuttonBarto2018}. The particular choice of $b = \pi_{\v \theta}$ (\textit{on-policy} learning) circumvents this issue, as the behavior policy presumably will be the best available estimate for the optimal policy at any given time. Although such a choice makes $b$ nonstationary, it is reasonable to expect that any undesired effects of this will only be transient as long as $\v \theta$ converges. 

Furthermore, like \cref{eq:pf_expectation}, \cref{eq:general_objective_gradient} is based on the assumption that $p(s_{i}) \propto w(s_{i}), i = 1,n$. If we were to restrict ourselves to the nonsymmetrized form of \cref{eq:general_objective_gradient}, $p(s_1) \propto w(s_1)$ could be achieved by drawing the initial states of the trajectories from an equilibrated MCMC simulation. If the symmetrized form is to be used, $p(s_{i}) \propto w(s_{i}), i = 1,n$ can only be guaranteed if the entire trajectories are taken from equilibrated MCMC simulations. Both approaches are clearly not desirable, since they run against the aim for an efficient, online policy optimization procedure. Specifically, if a trajectory $c = (s_1 \to s_2 \to \ldots \to s_n)$ is to be treated as a collection of subtrajectories, $\{(s_1 \to s_2 \to \ldots s_k), (s_2 \to s_3 \ldots s_{k+1}), \ldots \}$, then $p(s)$ for almost all $s \in c$ will depend on the particular dynamics of $b$, which is \emph{not} guaranteed to sample states proportionally to $w$. Nevertheless, we will ignore this fact and continue to treat the states \emph{as if} $p(s) \propto w(s)$. The bias introduced by doing so may at first seem severe, but note that a good behavior policy will only generate states close to the desired, true distribution of $w$; as the policy of the on-policy learning becomes better (due to optimization and/or improved modeling), the bias will diminish.

The above considerations taken into account, \cref{eq:general_objective_gradient} is approximated by
\begin{multline}
   \grad_{\v \theta} \ev{\epsilon_O}_{c \sim w} 
   \approx
  \frac{1}{2 N_c}\sum_{\substack{s_1 \sim \pi_{\v \theta} \\ c \sim \pi_{\v \theta}}} \epsilon_O(c) \e{-\frac{1}{2}\abs{\Delta f_{\v \theta}(c)}} \\
  \times \grad_{\v \theta} \left[\ln \pi_{\v \theta}(\overset{\to}{c}) + \ln \pi_{\v \theta}(\overset{\gets}{c}) - \abs{\Delta f_{\v \theta}(c)}\right],
  \label{eq:approximate_objective_gradient}
\end{multline}
where $N_c$ is the number of trajectories sampled. \Cref{eq:approximate_objective_gradient} is the form of the policy gradient we will be using in this work.

\subsection{Modeling the policy}\label{sec:modeling_the_policy}
There are very few restrictions on the functional form of a policy. As long as ergodicity and reversibility is preserved, the latter meaning that $\pi(s \to s')> 0 \Leftrightarrow \pi(s' \to s) > 0$, any function $\pi: \mathcal{A}\times S \to [0,1]$ will do. (Albeit only a few of those will do \emph{well}.)

Without lack of generality~\footnote{Strictly speaking, the functional form of \cref{eq:preference_softmax} forces $\pi(a|s) > 0 \forall a \in \mathcal{A}_s, s \in \mathcal{S}$, i.e. $\pi(a|s) = 0$ is now prohibited. In practice, however, there is no numerical difference between $\pi(a|s) = 0$ and $\pi(a|s) \to 0^+$ and the discrepancy can safely be neglected.}, we may be write
\begin{equation}
 \pi(a|s) = \frac{\exp[h(a|s)] }{\sum_{a' \in \mathcal{A}_s} \exp[h(a'|s)]} \in (0,1]
 \label{eq:preference_softmax}
\end{equation}
where $h(a|s)$ is the \textit{preference} for taking action $a$ from state $s$~\cite{SuttonBarto2018}. It is often computationally more convenient to work with the preference than the policy directly. One reason is that $h: \mathcal{A} \times \mathcal{S} \to \mathbb{R}$ is not restricted to a finite interval. Another is that the preference, being logarithmic, easily handles a large range of policy values without leading to numerical under or overflow.

In practice, the preference will be parametrized by $\v \theta$. It can be modeled by an arbitrary function approximator, although the PGMC performance will depend strongly on its ability to imitate an efficient Markov chain dynamics for the probability weight model in question -- and doing so efficiently: not only will \cref{eq:preference_softmax} have to be evaluated \emph{given} an action $a$, actions also have to selected \emph{based on} it, all while the state $s$ and possibly the action space $\mathcal{A}_s$ is dynamically changing during the simulation. Depending on the model to be simulated, a significant portion of the computer time may be spent on handling \cref{eq:preference_softmax}.

We will simply use linear \textit{feature maps} as preferences in this work~\cite{SuttonBarto2018}. Despite these having limited expressive power, they are fast, automatically incorporate translational invariance, scale efficiently, and are easy to deal with in the relatively simple scenarios explored in \cref{sec:PGMC_in_practice}. Utilizing more sophisticated constructs, like decision trees, neural networks, etc.,~\cite{Bishop:2006} is left for future works to explore.

Examples of more complex policy structures going beyond \cref{eq:preference_softmax} are presented in \cref{sec:PGMC_in_practice}.

\subsection{The performance factor}
The final element of the PGMC optimization to be discussed is the performance factor, as defined by \cref{eq:pf}.

The first factor of the denominator, the integrated autocorrelation time, is formally given by
\begin{align}
 \tau_{O} &= \lim_{\Delta \to \infty} \frac{1}{2}\sum_{\delta=-\Delta}^\Delta \rho_{O}(\delta), \label{eq:integrated_autocorrelation_time} \\
 \rho_{O}(\delta) &\equiv \frac{\ev{O_t O_{t+\delta}} - \ev{O}^2}{\ev{O^2} - \ev{O}^2}. \label{eq:rho}
\end{align}
for a trajectory of samples, $O_1 \to O_2 \to...$, where $O_i \equiv O(s_i)$. In practice, the series should be truncated when $\Delta$ is a few times larger than $\tau_{O}$ to give a meaningful result, since $\Var(\tau_{O}) \to \infty$ as $\Delta \to \infty$~\cite{Sokal1997}.

Estimating $\tau_{O}(c)$ based on \cref{eq:integrated_autocorrelation_time} is a viable strategy during the training stage if the training trajectories fed to the stochastic gradient decent optimization are only weakly correlated. If this is not the case, as suggested in \cref{sec:policy_search}, the change in $\tau_O(c)$ between two consecutive subtrajectories will be minuscule and sensitive to noise, and the optimization likely unstable. An alternative, heuristic approach is to simply use
\begin{equation}
 (2\tilde\tau_O)\inv \equiv 1 - \rho_O(1),
 \label{eq:tau_approximation}
\end{equation}
possessing the desired limiting properties of $(2\tilde\tau_O)\inv = 0$ for perfectly correlated samples and $(2\tilde\tau_O)\inv = 1$ for perfectly uncorrelated samples.

The take-home message of \cref{eq:tau_approximation} is that a rough estimate for $\tau_{O}$ can be obtained based only on pairs of consecutive samples, $(O_t, O_{t+1})$. Ultimately, this means that it should be viable to perform stochastic gradient decent with subtrajectories merely of length $2$. Obviously, one has to be careful when doing so, as such a ``myopic'' perspective may disregard important longer-time-scale dynamics; additional means may have to be employed to ensure that the policy converges to a reasonable target. When in doubt, the unbiased integrated autocorrelation time given by \cref{eq:integrated_autocorrelation_time} will always be the correct measuring rod to compare with. Nevertheless, and crucially from a practical perspective, the challenging problem of maximizing \cref{eq:pf_expectation} is now within reach of relatively simple program designs.

Among the many aspects that make up the cost factor $u$, only the cost associated with the policy is easily controllable and hence of interest here. The specifics are strongly case dependent, but two limiting regimes may be identified: I) If the cost of evaluating the model dominates a PGMC sampling iteration, the cost of evaluating the policy, i.e. selecting an action, is of minor importance and may be approximated by a (negligible) constant; in this case, the focus should be on getting the most out of each model evaluation, meaning that one should seek to minimize the integrated autocorrelation time in order to maximize \cref{eq:pf}. II) If, on the other hand, the cost of evaluating the model is small compared to that of the policy, the cost of the policy is obviously not negligible. In this case, a reasonable approximation would be to make $u(c)$ proportional to the number of operations needed to propose $c$.

Note that it should be sufficient to select a single, hopefully representative, slow mode (i.e. large autocorrelation time) observable for calculating the performance factor used in finding a good policy. Performing separate simulations for each observable of interest -- as opposed to sample all of them during the same simulation -- is most likely not an efficient strategy.

\section{PGMC in practice}\label{sec:PGMC_in_practice}
In this section, we will demonstrate how the PGMC framework can be employed in simulating a simple, but challenging lattice model. The solutions presented are not necessarily meant to be the best ones, but are selected to gradually expose some of PGMCs potential. It should also be emphasized that PGMC is a general scheme and not restricted to the examples presented here. To underline this, we briefly share some general thoughts on the usage of PGMC in other cases at the end of the section, in \cref{sec:other_systems}.

See \cref{app:simulation_details} for technical details about the simulations.

\subsection{Test model}\label{sec:test_model}
\begin{figure}
 \includegraphics[width=1.0\columnwidth]{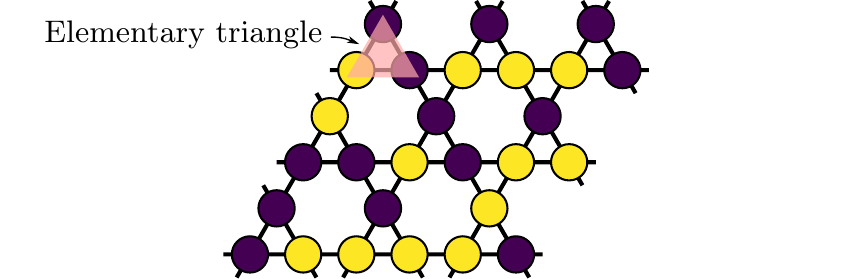}
 \caption{A configuration of the Ising model on a $L=3$ kagome lattice. The dark and light disks indicate down (-1) and up (+1) spins, respectively.}
 \label{fig:kagome_lattice}
\end{figure}

\begin{figure}
 \includegraphics[width=1.0\columnwidth]{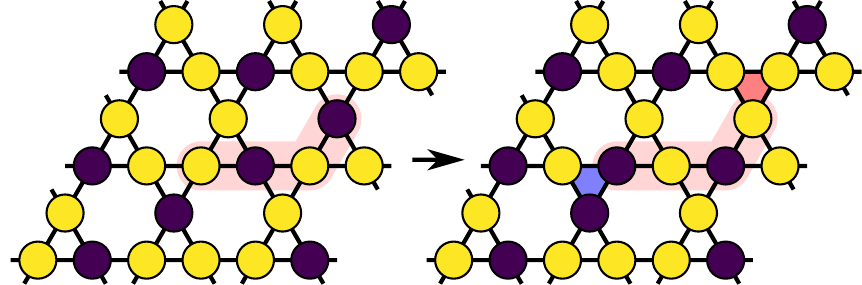}
 \caption{Generation of two separated monopole excitations from a kagome spin ice ground state by flipping a string of alternating spins. The monopoles are associated with the colored elementary triangles -- the only ones that violate the spin ice constraint by not containing two up spins (light) and one down spin (dark).}
 \label{fig:spin_ice_monopoles}
\end{figure}

\begin{figure}
 \includegraphics[width=1.0\columnwidth]{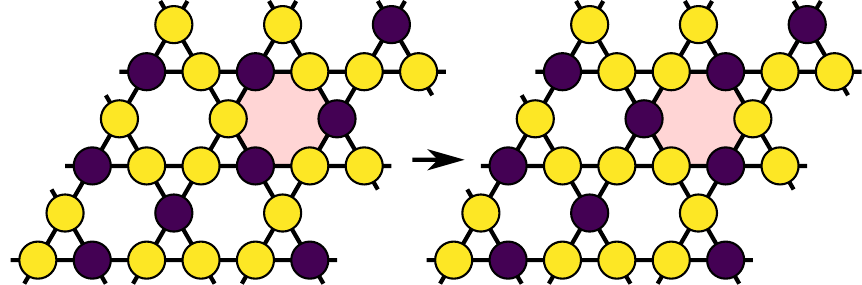}
 \caption{Tunneling from one kagome spin ice ground state to another. A loop of spins of alternating signs has to be flipped in order to conserve the spin ice rule everywhere. The smallest loop for which this can be accomplished belongs to an elementary hexagon or \textit{plaquette}.}
 \label{fig:spin_ice_tunneling}
\end{figure}

As a test model, we choose the Ising model on a kagome lattice of size $N = 3 \times L^2$ with periodic boundary conditions; see \cref{fig:kagome_lattice}. 
Now, the state space is $\mathcal{S} = \{\v \sigma | \sigma_i \in \pm 1\}$ and the log probability weight is given by
\begin{equation}
\ln w(\v \sigma) = K \sum_{\nn{i,j}} \sigma_i \sigma_j - K B \sum_i \sigma_i,
\label{eq:Ising_model}
\end{equation}
with $K$ being a coupling constant and $B$ the strength of an external field. This simple model displays a surprisingly rich physics: In the thermodynamic limit, for $0<K<K_\text{c}$ and $B=0$, it is a paramagnet, which at $K_\text{c} = \frac{1}{4}\ln(3+\sqrt{12}) \approx \num{0.467}$ undergoes a continuous phase transition to a ferromagnetically ordered state~\cite{Syozi:1951}. Conversely, when $K<0$, the model is frustrated. As $K \to -\infty, B=0$, it does not experience a phase transition, but rather a crossover to an extensively degenerate ground macrostate~\footnote{We use the term \textit{macrostate}, instead of the more customary \textit{state}, to explicitly distinguish this physical mode from the \textit{microstate} meaning attached to the word \textit{state} elsewhere.} with a residual entropy of $\approx 0.502$ per spin~\cite{KanoNaya:1953}. In the ground macrostate, the elementary triangles are constrained to contain either two up and one down spin, or one up and two down spins. The configurations belonging to this \textit{manifold} of states are connected by a series of single spin flips, each of zero energy cost. Applying a field of strength $0 < \abs{B} < 4$ partially lifts the degeneracy, since now only one of the elementary triangle configuration categories will minimize the energy. This reduces the residual entropy to $\approx 0.108$ per spin, and the system enters the \textit{kagome spin ice} submanifold -- so named for its connection to (pyrochlore) spin ice~\cite{UdagawaOgataHiroi:2002}. This macrostate is topologically nontrivial, with particle-like, gapped excitations dubbed \textit{monopoles}~\cite{MacdonaldETAL:2011,CastelnovoETAL:2008}. See \cref{fig:spin_ice_monopoles}. 
In order to tunnel from one spin ice configuration to another, a minimum of six spins in a loop has to be flipped, as illustrated in \cref{fig:spin_ice_tunneling}.

The simple model of \cref{eq:Ising_model} provides a rich test bed for PGMC algorithms, whereby the intrinsic dynamics, and hence the nature and ``difficulty'' of the simulation, can be tuned by only two parameters. In the following, we will consider the representative set of parameters given in \cref{tab:parameters} for benchmarking. 

\begin{table}
  \caption{\label{tab:parameters}Parameters used in the simulations of \cref{eq:Ising_model}. The parameters have been chosen as to realize the various physical -- hence also simulation -- regimes of the model.}
  \begin{ruledtabular}
  \begin{tabular}{ddl}
  K & B & Description \\
  \colrule
  K_\text{c} \approx 0.467 & 0 & The critical point. \\
  0.5 & 1 & Ferromagnetic and nonzero field. \\
  -4 & 0 & Strong frustration and zero field. \\
  -4 & 0.5 & Strong frustration and weak field. \\
  -4 & 2.5 & Strong frustration and strong field. \\
  \end{tabular}
  \end{ruledtabular}
\end{table}

\subsection{Single flip policies}
To demonstrate the basics of PGMC, we compare the performance of a few simple, but increasingly complex policy models. The policies do all have action spaces restricted to single site spin flips, 
\begin{gather}
 \mathcal{A} = \bigl\{a_{\sigma_i} \bigm| i \in \{1,\ldots,N\}\bigr\}, \\
 a_{\sigma_i} : \sigma_j \mapsto 
  \begin{cases}
    -\sigma_j, & j = i \\
    \hphantom{-}\sigma_j, & j \ne i
  \end{cases}, \quad a_{\sigma}\inv = a_{\sigma}
\end{gather}
the minimum required for ergodicity to be possible. They are:

\begin{enumerate}
 \item[$\pi_1$.] The \textit{trivial policy}, i.e. uniformly random spin selection as given by preference
 \begin{equation}
  h_1(a_{\sigma_i}|\v \sigma) = \frac{1}{N}.
 \end{equation}
 This corresponds to a canonical Metropolis MCMC, with no independent parameters to be determined.
 \item[$\pi_2$.] The \textit{local spin orientation policy}, given by
 \begin{equation}
  h_2(a_{\sigma_i}|\v \sigma) = 
    \begin{cases}
      \theta_1 & \text{if } \sigma_i = 1 \\
      \theta_2 & \text{if } \sigma_i = -1
  \end{cases}.
 \end{equation}
 With $\pi_2$, the probability of flipping a spin may depend on its sign. Although we expect $\theta_1 = \theta_2$ to be the right choice in zero external field, a distinction may be advantageous when the global $\mathbb{Z}_2$ symmetry of \cref{eq:Ising_model} is explicitly broken. The cost is that one independent parameter has to be tuned.
 \item[$\pi_3$.] The \textit{local energy sign policy}, with preference
  \begin{equation}
  h_3(a_{\sigma_i}|\v \sigma) = 
    \begin{cases}
      \theta_1 & \text{if } E_{\text{local},i} > 0 \\
      \theta_2 & \text{if } E_{\text{local},i} \le 0
      \label{eq:h_3}
  \end{cases},
 \end{equation}
 where $E_{\text{local},i} \equiv \sigma_i(\sum_{\nn{j,i}}\sigma_j - B)$, the sum being over the nearest neighbor sites of $i$. It is quite typical that the physics of a model is determined by only a few excited degrees of freedom, surrounded by a ``sea'' of less relevant, relaxed ones. This means that the trivial policy $\pi_1$ may ``waste'' most of its action proposals on unimportant -- and mostly rejected -- update attempts. $\pi_3$ is designed to combat this by making a distinction in selecting (locally) excited and non-excited spins. Compared to $\pi_2$, $\pi_3$ takes more information into account, rendering it more versatile and potent. This comes with the expense of more complexity. As with $\pi_2$, there is now one independent parameter to be determined.
 \item[$\pi_4$.] The \textit{local mean field policy}, with 
 \begin{equation}
  h_4(a_{\sigma_i}|\v \sigma) = \theta_{q_i},
  \label{eq:h_4}
 \end{equation}
 where
 \begin{equation}
   q_i = \frac{1}{2}\left[(\sigma_i + 1) + 2\sum_{\nn{j,i}}(\sigma_j + 1)\right] + 1
   \label{eq:feature_map}
 \end{equation}
 associates a unique integer $\in \{1,2,\ldots,2(z+1)\}$ to each possible pair of $(\sigma_i, \sum_{\nn{j,i}}\sigma_j)$. For the kagome lattice, the coordination number is $z = 4$, so there are now 9 independent parameters to be fixed. $\pi_4$ contains all of the other policies, in the sense that $\pi_1$, $\pi_2$, and $\pi_3$ may be considered instances of $\pi_4$ with a restricted parameter space. It is therefore guaranteed to perform at least as well as them, given sufficient training~\footnote{Likewise, no linear combination of the form $\tilde \pi = \sum_{\alpha = 1}^4 \omega_i \pi_i$ for some $\omega_i > 0$ can surpass the performance of just $\pi_4$; $\tilde \pi$ will have the exact same ability to categorize local spin configurations as $\pi_4$}.
\end{enumerate}
We have omitted the subscript $\v \theta$'s to avoid clutter in the above expressions. Like the model, \cref{eq:Ising_model}, the policies $\pi_1$--$\pi_4$ are translational invariant. However, except trivially for $\pi_1$, they do not take advantage of the global $\mathbb{Z}_2$ symmetry when $B=0$.

The parameters $\v \theta$ are determined during the learning stage in an on-policy fashion, as outlined in \cref{sec:policy_search}.
Setting the autocorrelation observable to 
\begin{equation}
 O(\v \sigma) = \v \sigma,
 \label{eq:identity_observable}
\end{equation}
while treating the spin state as a vector with scalar product as multiplication, the performance factor may be approximated by just $\tilde \epsilon_O(c) = \text{constant}$. This is because all available actions will change the state by exactly the same amount, namely one spin flip, meaning that \cref{eq:tau_approximation} will be constant during the policy search. We also assume that the costs of evaluating the single flip policies are all similar. Then, if only a single two-state subtrajectory $c_t = (s_{t} \to s_{t+1})$ is kept in memory, \cref{eq:gradient_log_pf} reduces to
\begin{equation}
 \grad_{\v \theta} \left[\ln \pi_{\v \theta}(\overset{\to}{c}_t) + \ln \pi_{\v \theta}(\overset{\gets}{c}_t) - \abs{\Delta f_{\v \theta}(c_t)}\right],
 \label{eq:minimal_constant_pf_gradient}
\end{equation}
which is simply the gradient of the log-likelihood of creating this subtrajectory.

\begin{figure}
 \includegraphics[width=\columnwidth]{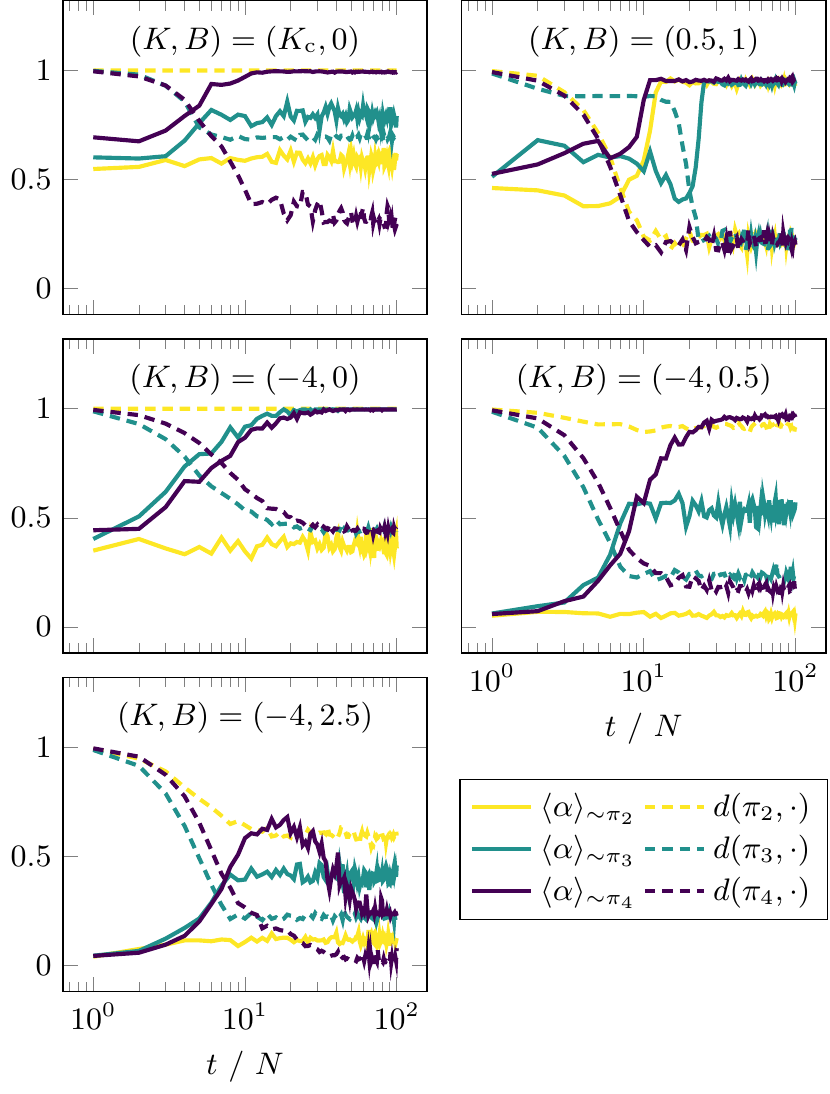}
 \caption{Learning curves for the nontrivial single flip policies $\pi_2$, $\pi_3$, and $\pi_4$, (with $\pi_x = \pi_x(t)$) showing the expected acceptance rate $\ev{\alpha}$ and normalized effective degrees of freedom $d$ as a function of optimization iterations. The quantities displayed are averages over sweeps of $N$ optimization iterations. Each case was initialized with a random configuration and $\v \theta = \v 0$.
 }
 \label{fig:single_flip_learning_curves}
\end{figure}

Examples of the policy optimization, as monitored by the mean acceptance rate $\ev{\alpha}$ and the normalized effective degrees of freedom~\footnote{The effective degrees of freedom may also be seen as the exponential of the Shannon entropy~\cite{Shannon:1948} of the policy for a given state; it basically measures the amount of information encoded in the policy, with a lower entropy corresponding to less uncertainty of which action that will be selected next.},
\begin{equation}
 d(\pi, s) \equiv \frac{1}{N}\exp(-\sum_{a \in \mathcal{A}_s} \pi(a|s)\ln \pi(a|s)),
 \label{eq:effective_dof}
\end{equation}
are shown in \cref{fig:single_flip_learning_curves}. (Note that in this case, since $\tilde \epsilon_O = \text{constant}$ and $b = \pi_{\v \theta}$, $\ev{\tilde \epsilon_O} \propto \ev{\alpha}$. If this does not hold, it is more reasonable to track $\ev{\tilde \epsilon_O}$ directly.)
After a short initial relaxation phase, the policies quickly converge to ``focus'' on a subset of more relevant degrees of freedom, with $\pi_4$ focusing more than $\pi_3$, and $\pi_3$ more than $\pi_2$. Concurrently, this leads to a bias towards sampling physically relevant states, which in turn makes a more refined tuning of the parameters possible. The overall effect is to increase the acceptance rate and decrease the effective number of degrees of freedom, with the additional benefit of improving the equilibration of the system. As expected, $\pi_2$ is not able to \textit{learn} anything from the $B=0$, non-symmetry broken cases. The increase in number of parameters from $\pi_2$ and $\pi_3$ to $\pi_4$ means that the convergence typically -- but not always --  is slower for the latter: The number of iterations required increases because finding a convergent point in a higher dimensional parameter space requires more information. On the other hand, an improved knowledge about the dynamics may also help with a quicker relaxation of the system, somewhat countering, and in the case of $(K,B) = (0.5,1)$, exceeding, this effect. The peculiar, nonmonotonic behavior of $\ev{\alpha}_{\sim \pi_4}$ for $(K,B) = (-4,2.5)$ most likely relates to the challenging state space in this regime: At this point in $(K,B)$ space, the model is close to the kagome spin ice ground macrostate. The remaining energy above the ground macrostate is associated with localized monopoles and can only be released by their mutual annihilation. Hence, while the monopoles diffuse around ``in search of an annihilation partner'', $\pi_4$ is able to temporarily learn the ``incorrect'' dynamics of these excitations, before the ground macrostate is finally found and the parameters are adjusted accordingly. $\pi_2$ and $\pi_3$, on the other hand, do not have the capacity to model the physical macrostate properly, and do therefore converge to a less restricted subset of state space.

\begin{figure}
 \includegraphics[width=\columnwidth]{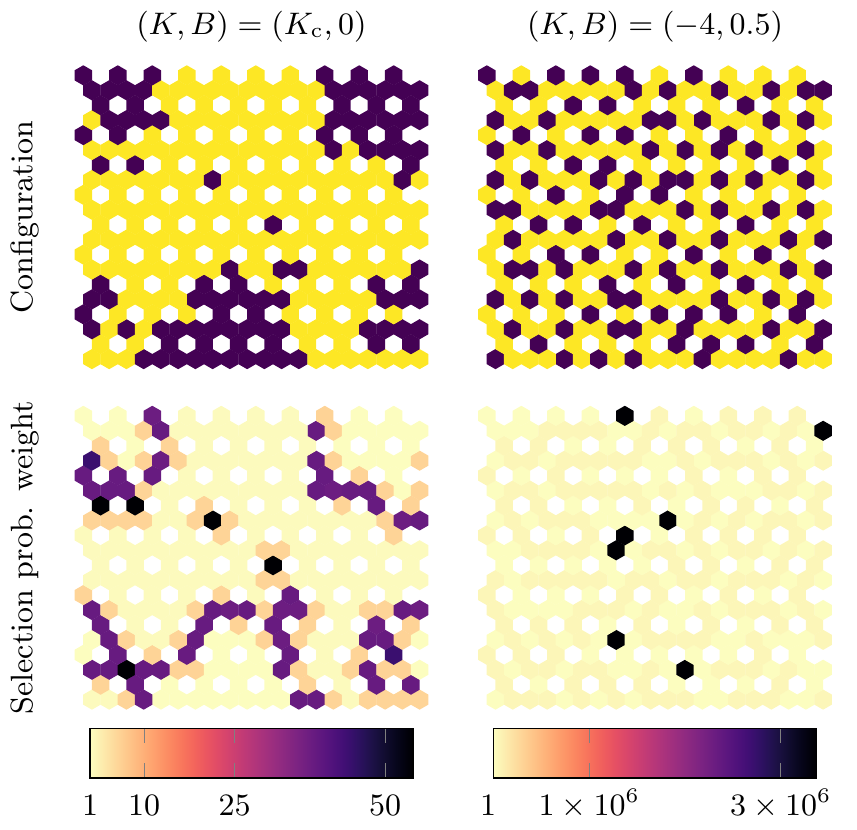}
 \caption{Snapshots of configurations (top) and the corresponding relative probability weights for selecting the spin to flip in the next iteration (bottom), according to policy $\pi_4$. The state on the left is sampled at criticality, while the state on the right is frustrated. The policy, separately optimized to each parameter point, assigns a larger weight to energetic structures like domain walls and monopole excitations, and relatively more so in the case of the frustrated state close to the ground macrostate. The connectivity lines of \cref{fig:kagome_lattice,fig:spin_ice_tunneling} have been omitted, and only the values associated with the sites (the small hexagons) are shown.
 }
 \label{fig:single_flip_PGMC_snapshots}
\end{figure}

The policies ability to automatically learn to focus on important degrees of freedom may also be visualized directly, as is done in \cref{fig:single_flip_PGMC_snapshots}. 

\begin{figure}
 \includegraphics[width=0.97\columnwidth]{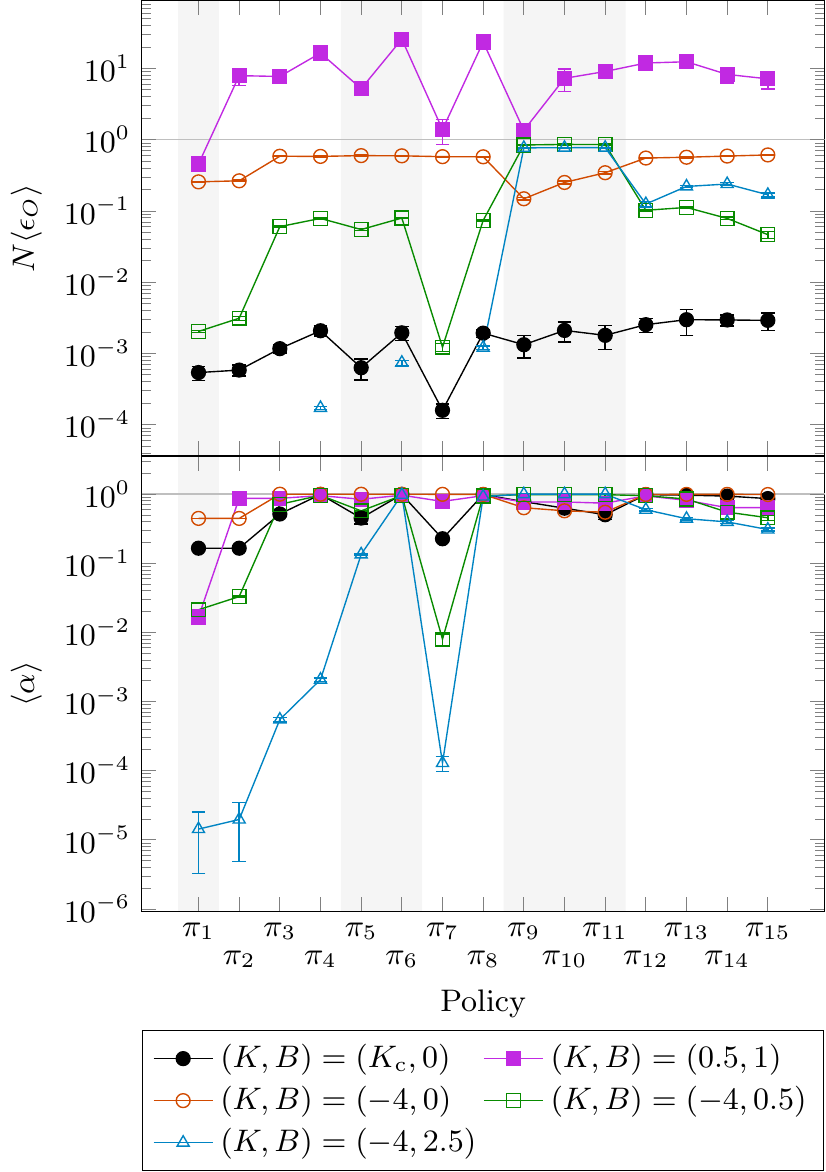}
 \caption{
 PGMC sampling performance of \cref{eq:Ising_model} for the parameters listed in \cref{tab:parameters} and all policies investigated in this work, after being optimized. The upper plot shows the estimated performance factor, \cref{eq:pf}, calculated from \cref{eq:identity_observable,eq:integrated_autocorrelation_time} and the cost factor $u(c) = $ \textit{number of elementary actions used in constructing $c$}. The performance factor has been scaled by the number of sites $N$ to make the time scale correspond to a \textit{Monte Carlo sweep}. The lower plot shows the acceptance rate. The horizontal lines indicate the ``optimal'' performance, i.e. what the results would have been had the configurations been drawn directly following \cref{eq:Ising_model}, instead of being generated by Markov chains. The background shading indicate classes of policies sharing a similar structure. Results obtained at the same $(K,B)$ point share a common color and symbol style, with lines as guides to the eye. Results in unfrustrated parameter regimes are shown with filled symbols, while the frustrated ones are shown with open symbols. In some cases, for $\pi_1$--$\pi_3$, $\pi_5$ and $\pi_7$, finding $\ev{\epsilon_O}$ for $(K,B) = (-4,2.5)$ was too slow/unreliable, so these points have been omitted. Error bars indicate sample standard deviation. See the main text and \cref{app:simulation_details} for discussion and further details.
 }
 \label{fig:MCMC_performance}
\end{figure}

After convergence, the learned policies are used in the second stage of PGMC, the actual MCMC simulation. Their relative performances are evaluated by comparing estimates of $\ev{\epsilon_O}$ -- obtained from \cref{eq:pf,eq:integrated_autocorrelation_time,eq:identity_observable}. The cost factors are set to be equal to the number of \textit{elementary actions} -- i.e. single spin flips -- involved in a PGMC step, which in these cases equals one. The results are summarized in \cref{fig:MCMC_performance}, where the measured acceptance rates are also plotted.
The trends of the training results repeat here: The performance typically increases, sometimes significantly, by going from $\pi_1$ to $\pi_4$, with the notable and expected exception of $\pi_2$ when $B=0$. Sampling at $(K,B) = (-4,-2.5)$ is by far the most challenging task, for which all the single flip policies, except maybe $\pi_4$, fails. Tracking the performance at this parameter point will therefore be of great interest later on. The seemingly low yield at the critical point is a consequence of measuring the performance using the observable of \cref{eq:identity_observable}, which does not take the global $\mathbb{Z}_2$ symmetry into account; the autocorrelation length is therefore determined by the long time scale of reversing the global polarization. Another interesting manifestation of the focusing ability of the (nontrivial) policies appears at the ferromagnetic and polarized $(K,B) = (0.5,1)$. As $N \ev{\epsilon_O} > 1$ in this case, the PGMC simulations are able to \emph{outperform} a hypothetical ``perfect sampling'' of drawing directly from the desired probability distribution. The reason is straightforward: As the physically relevant states are close to being completely polarized, the effective number of degrees of freedom is significantly lower than the actual number of degrees of freedom ($N$). Most spins are left untouched during an update, which corresponds to ``updating'' them ``for free''. On the contrary, constructing a configuration from scratch, however efficiently it is done, comes with a cost proportional to the actual number of degrees of freedom. At some point, the former starts outperforming the latter.

For several of the parameter choices, especially when using $\pi_4$, the PGMC simulations have close to 100\% acceptance rates, i.e. they are almost rejection free. This means that the policies are able to capture almost the entire Markov chain dynamics, similar to what is achieved by algorithms hand crafted to do so (see, for example, Refs. \cite{Bortz_et_al:1975,FichthornWeinberg:1991,PhysRevE.85.026703}). This does \emph{not} mean that these policies are optimal, in the sense of maximizing $\ev{\epsilon_O}$, but rather that the updates proposed by them, however correlated, closely follows the desired probability distribution of \cref{eq:Ising_model}.

\subsection{Chain policies}
So far, we have encountered just a small fraction of the vast space of possible policies. The single flip policies $\pi_1$--$\pi_4$ do only -- at most --  take information about nearest neighbor spins into account. This is not always sufficient for a satisfying PGMC sampling, as is exemplified by the challenging parameter point of $(K,B) = (-4,2.5)$. 

There are several possible directions to explore in a quest for improving this. One would be to increase the amount of information ``available'' for each action selection by taking further neighbors into account. It is easy to imagine that this could help a policy in guiding the Markov chain more precisely and hence improve the sampling performance. However, as a single spin flip policy can only change a configuration by one spin flip at a time, such an approach (alone) would still struggle with state space \textit{probability barriers} of several spin flips.

A more radical way is to enlarge the action space by allowing for more spin flips between the Metropolis acceptance steps. In other words, let an action consist of several elementary actions instead of just one. Doing so increases the entropy associated with selecting an action, making longer jumps in state space possible and freezing of the Markov chain less likely.

\begin{figure}
  \includegraphics[width=0.8\columnwidth]{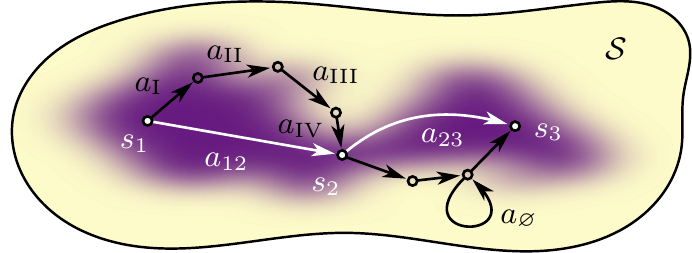}
  \caption{Example of a three state trajectory of a chain policy of length four. Each action, $a_{12}: s_1 \mapsto s_2$ and $a_{23}: s_2 \mapsto s_3$, consists of four elementary actions. The Metropolis acceptance step is only concerned with the states before and after the actions have been completed (dots with white fill), making room for temporary visits of transit states (empty dots) with lower probability weights.}
 \label{fig:chain_policy}
\end{figure}

We will pursue the latter approach here, by what we call \textit{chain policies}. A chain policy consists of a number of \textit{elementary policies}, each successively selecting and applying an elementary action to the most recent (\textit{transit}) state, as illustrated in \cref{fig:chain_policy}. For example, a chain policy $\pi$ of the two elementary policies $\pi_\text{I}$ and $\pi_\text{II}$ is given by
\begin{equation}
 \pi(s\to s') = \pi_\text{I}(s \to s'')\pi_\text{II}(s'' \to s'),
 \label{eq:length_two_chain_policy}
\end{equation}
where the action $a: s \mapsto s'$ is composed of the two elementary actions $a_\text{I}: s \mapsto s''$ and $a_\text{II}: s'' \mapsto s' $, i.e. $a = a_\text{II} a_\text{I}$. Using $\pi$, the probability weight of the transit state $s''$ will not be subject to the Metropolis acceptance step, \cref{eq:metropolis_acceptance}, opening up for possibility of tunneling probability barriers. $s''$ is still needed in calculating the policy of the inverse process, which simply reads $\pi(s' \to s) = \pi_\text{I}(s' \to s'')\pi_\text{II}(s'' \to s)$. Note that the elementary policy of going from one (transit) state to the next does \emph{not} need to be the same for the inverse action (compare $\pi_\text{I}(s \to s'')$ with $\pi_\text{II}(s'' \to s)$). 

Now, if a chain policy is to be constructed based on the single flip policies of the previous subsection, the Markov chain will no longer be ergodic: In the above example of a chain policy of length two, there is no way in which a configuration with an odd number of spins pointing up can be reached from a configuration with an even number of spins pointing up -- a consequence of the binary nature of the spins~\footnote{Flipping two spins leads to an overall change of $\pm 1 \pm 1 = 0,\pm 2$ in the number of spins pointing up. Adding an even number to an even(odd) number is an even(odd) number}. Therefore, to ensure ergodicity, we are forced to extend the action space of the single flip policies with a \textit{do-nothing} action, 
\begin{equation}
a_\varnothing: s \mapsto s, \quad a_\varnothing\inv = a_\varnothing, 
\end{equation}
and assign an additional preference weight of $\theta_\varnothing$ for selecting this~\footnote{Alternatively, we can extend the elementary action space to encompass \emph{both} single spin flips \emph{and} multiple spin flips. Such a solution is however significantly more complicated than extending it with a do-nothing action. That said, the main point is that the policy has to be made ergodic. Should the degrees of freedom of a model to be simulated be of a different character, e.g. continuous instead of binary, the elementary policies may not need to be amended when used in a chain policy}.

The increased complexity of the action space makes the policy training more complicated as well. The actions may now result in a net $n_\text{flips}(s \to s') \in \{0,\ldots, n\}$ spin flips when updating $s \to s'$ with a length $n$ chain policy. (In this regard, a \textit{bounce} process, $a_{\sigma}$ followed by $a_{\sigma}\inv$, constitutes zero spin flips if it happens within an update, since $a_{\sigma}\inv a_{\sigma} = a_{\varnothing}$.) For simplicity, we model the performance factor for the chain policies as
\begin{equation}
 \tilde \epsilon_O(s \to s') = n_\text{flips}(s \to s').
 \label{eq:chain_policy_pf}
\end{equation}
This should, to leading order, be proportional to the true value for policy chains of moderate lengths, again assuming that the elementary policies all have similar cost factors. Since \cref{eq:chain_policy_pf} is not constant, \cref{eq:minimal_constant_pf_gradient} can no longer be used. Instead, a minimum of three states, $s_{t-2} \to s_{t-1} \to s_t$, has to be memorized, such that two subtrajectories of lenght 2, $c_1 = (s_{t-2} \to s_{t-1})$ and $c_2 = (s_{t-1} \to s_t)$, can be used in approximating $\ev{\epsilon_O}$. This is necessary if \cref{eq:gradient_log_pf} is to be calculated without the performance factors in the denominator and numerator canceling~\footnote{In case $\tilde\epsilon_O(c_1) = \tilde\epsilon_O(c_2) = 0$, \cref{eq:gradient_log_pf} can, for instance, be set to zero. This is what has been done in this work}.

We first test the following two chain policies of length 2:
\begin{itemize}
 \item[$\pi_{5}$.] Elementary preference like $h_3$, \cref{eq:h_3}, with the addition of $a_\varnothing$ to the elementary action space. This increases the number of independent parameters to 2.
 \item[$\pi_{6}$.] Elementary preference like $h_4$, \cref{eq:h_4}, with the addition of $a_\varnothing$ to the elementary action space. This increases the number of independent parameters to 10.
\end{itemize}
For simplicity, we have kept the elementary policies of these chain policies equal (i.e. $\pi_\text{I} = \pi_\text{II}$). 

The performance of PGMC sampling with $\pi_5$ and $\pi_6$, after training, is also shown in \cref{fig:MCMC_performance}. Some trends are clear: The inclusion of $a_\varnothing$ in the action space means that the Markov chain dynamics may now follow the probability distributions of \cref{eq:Ising_model} much closer in all parameter regimes, as the Metropolis rejections are modeled as well. This leads to a very high acceptance rate, especially for $\pi_6$, which has the most flexible elementary policy. Were we to simulate a model that is expensive to evaluate, this feature may be of significance in itself: it is redundant to calculate the probability weight in the Metropolis acceptance step if it is known, in advance, that the state will not change. In the current case, however, such a saving is negligible. For the field free parameter points, $(K,B) = (K_{\text{c}},0)$ and $(K,B) = (-4,0)$, the performance factor estimates do not change much as compared to the results of the corresponding single flip policies; from the perspective of the Markov chain dynamics, there is little difference in taking one or two elementary action steps at a time. If anything, there seems to be a disadvantage for $\pi_5$ at the critical point, as the imperfect tracking of the desired probability distribution by the elementary policy is enhanced when not corrected before after two iterations. The situation is more interesting with stronger fields. There is an additional energy cost of $2B$ associated with flipping a single spin, i.e. there is a probability barrier of exciting a spin against a nonzero field. The reverse operation can not always counter this, as the gain of $-2B$ may be ``more than necessary'' for Metropolis acceptance. However, by combining two flips, one of a spin parallel and one of a spin antiparallel to the field, the overall energy change associated with the field cancels; the barrier has been tunneled, or at least lowered. This effect is readily seen for $\pi_6$, with an improvement over $\pi_4$ for both $(K,B) = (0.5,1)$ and $(K,B) = (-4,2.5)$. Similar gains are expected for $\pi_5$ over $\pi_3$ as well, but like mentioned above, the benefit seems to be outweighed by the cost of insufficient tracking of the probability distribution.

It is tempting to repeat the exercises of $\pi_5$ and $\pi_6$, but with chain policies 
\begin{itemize}
 \item[$\pi_7$.] Like $\pi_5$, but with length 6.
 \item[$\pi_8$.] Like $\pi_6$, but with length 6.
\end{itemize}
The motivation is that this opens up for a low, but nonzero probability of direct tunneling between kagome ice ground states [\cref{sec:test_model}]. In order to promote such processes, we skew the training performance factor estimate from \cref{eq:chain_policy_pf} to
\begin{equation}
 \tilde\epsilon_O(s\to s') = \max[0, n_{\text{flips}}(s\to s')-2],
 \label{eq:skewed_chain_policy_pf}
\end{equation}
hence ignoring training samples that involve two or fewer spin flips. 

\Cref{fig:MCMC_performance} shows that $\pi_7$ and $\pi_8$ continue the trends of $\pi_5$ and $\pi_6$: When the elementary policy is able to follow the probability distribution given by the model, \cref{eq:Ising_model}, the performance is similar or improved, while if not, the ``punishment'' is more severe. 

Observe that $\pi_7$ and $\pi_8$ are the only policies presented so far that have the potential to properly sample the true kagome spin ice ground macrostate, $K \to -\infty, B \ne 0$ -- regardless of available computer power.

\subsection{Stochastic chain policies}\label{sec:stochastic_chain_policies}
\begin{figure}
 \includegraphics[width=0.85\columnwidth]{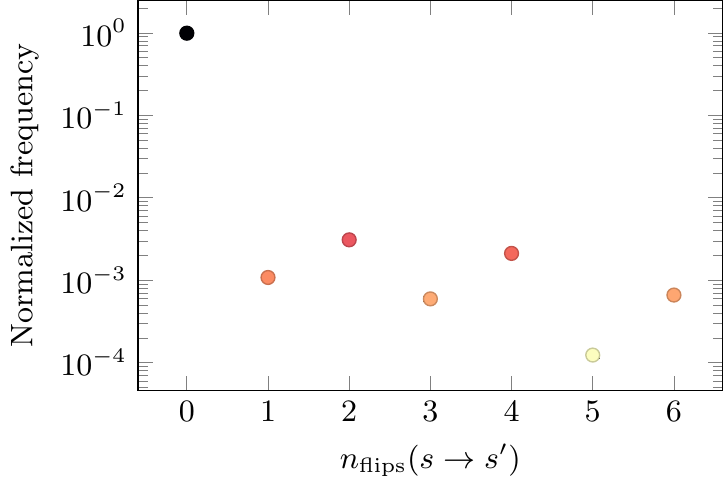}
 \caption{The frequency of updates leading to $n_\text{flips}(s \to s') \in \{0,\ldots,6\}$ net spin flips in a PGMC sampling of \cref{eq:Ising_model} at $(K,B) = (-4,2.5)$ with policy $\pi_8$. Most updates involve no change, either because the proposed action effectively is $a_\varnothing$, or because the proposed update was rejected. The error bars, indicating the sampled standard deviation, are smaller than the symbol size. See \cref{app:simulation_details} for technical details.}
 \label{fig:spins_flipped_frequency}
\end{figure}

\Cref{fig:spins_flipped_frequency} reveals what can also be deducted from \cref{fig:MCMC_performance}: At $(K,B) = (-4,2.5)$, most PGMC iterations guided by $\pi_8$ do not lead to a change in configuration. Furthermore, most of the spin flips that do take place are basically spent at ``undoing'' previous flips. Overall, the Markov chain diffusion through state space is computationally inefficient and very slow.

\begin{figure}
  \includegraphics[width=0.8\columnwidth]{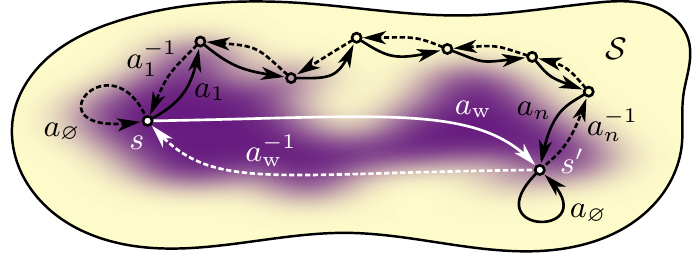}
  \caption{A worm action $a_\text{w} = a_\varnothing a_n\cdots a_1 : s\mapsto s'$ (solid arrows) and its inverse $a_\text{w}\inv = a_\varnothing a_1\inv \cdots a_n \inv: s'\mapsto s$ (dashed arrows). }
 \label{fig:worm_policy}
\end{figure}

Loop and worm algorithms~\cite{PhysRevLett.65.941,PhysRevE.57.1155,PhysRevB.82.134420,PhysRevE.85.036704,PhysRevB.90.220406,PhysRevE.96.023304}, hand crafted for taking advantage of the stringy structure of the ground states of many frustrated models, like \cref{eq:Ising_model} with $K\ll 0$, will typically perform much better than the policies used so far. This raises the question of whether it is possible to construct a policy model that can mimic such an algorithm. The answer is affirmative, as can be demonstrated with the following \textit{worm policy} scheme~\footnote{Only one out of several possible worm policies is given; others may be constructed by permuting the order in which the movement, flip, and termination actions are selected. Note also that the worm policy operates in the ordinary spin space, as opposed to the dual space representation used in e.g. Ref.~\cite{PhysRevE.96.023304}}:
\begin{enumerate}
 \item Initially, select and flip a random spin at site $i_1$ according to the single flip policy $\pi_\text{start}$, $a_{1} = a_{\sigma_{i_1}}$. (This time it is not necessary to include $a_{\varnothing}$ in the elementary action space.) Mark the site as the \textit{head of the worm}.
 \item For $t=2,3,\ldots$, until the worm construction is terminated after $n$ spin flips: According to elementary policy $\pi_\text{move}$, \emph{either} pick a nearest neighbor site $j$ of the head $i_{t-1}$, flip $\sigma_j$ ($a_{t} = a_{\sigma_j}$), move the head $i_{t} \gets j$ and repeat the step with $t \gets t+1$, \emph{or} terminate the worm construction ($a_{t} = a_{n+1} = a_\varnothing$). 
 \item The complete worm construction, $s = s_1 \to a_{\text{w}}s \equiv a_\varnothing a_n\cdots a_1 s_1 = s_{n+1} = s'$, can then be seen as the action chosen by the worm policy, 
 \begin{equation}
 \begin{split}
  \pi_{\text{worm}}(s \to s') = {} & \pi_\text{start}(s_1 \to s_2) \\
  & \times \left[\prod_{t = 2}^n \pi_\text{move}(s_t \to s_{t+1})\right] \\ 
  & \times \pi_\text{move}(s_{n+1} \to s_{n+1}),
 \end{split}
 \end{equation}
 where $s_{t+1} = a_t s_t$.
\end{enumerate}
Observe that the inverse action $a_{\text{w}}\inv$ will \emph{not} follow the exact same chain of elementary actions as $a_{\text{w}}$ (in reverse), since, in order for the worm policy to be reversible, $a_{\text{w}}\inv = a_\varnothing a_1\inv \cdots a_n\inv \ne a_1\inv \cdots a_n\inv a_\varnothing$. See \cref{fig:worm_policy}. 

The worm policy is an example of a chain policy of stochastic length, or \textit{stochastic chain policy}. In a stochastic chain policy, a good distribution of lengths will be (indirectly) machine learned, instead of fixed a priori by a \textit{hyperparameter}. 

The training of the stochastic chain policy proceeds in very much the same fashion as before, but now with the training performance factor estimate, \cref{eq:skewed_chain_policy_pf}, normalized by the length of the worm, $n$. It is important to incorporate such a normalization to reflect the cost of creating a longer worm and prevent the average worm length from diverging.

We test the above sketched worm policy in the form of
\begin{itemize}
 \item[$\pi_9$.] $\pi_\text{start} = \pi_4$; the selection of the initial spin flip is modeled exactly as $\pi_4$. $\pi_\text{move}$ is given by the same preference, \cref{eq:h_4}, (with a different set of parameters $\v \theta$) but with the action space $\mathcal{A}_{\v \sigma, i} = a_\varnothing \cup \left\{a_{\sigma_j} \mid j \text{ n.n. of $i$} \right\}$, where $i$ is the current position of the head. In total, there are $9+10 = 19$ independent parameters to be determined.
\end{itemize}
\Cref{fig:MCMC_performance} reveals $\pi_9$'s success in simulating the kagome ice regime, with close to ``perfect'' sampling at $(K,B) = (-4,0.5)$ and $(K,B) =(-4,2.5)$, orders of magnitude better than what is achieved by $\pi_1$--$\pi_8$. In the other parameter regimes, where there important states are of another, less stringy character, the worm policy does not offer any advantages.

The universality and great flexibility of the PGMC framework makes it easy to explore more sophisticated policy models. For example, it is only a matter of imposing a few restrictions on the action space of the worm policy to essentially mimic -- and extend -- the highly specialized directed worm algorithms~\cite{PhysRevE.66.046701,PhysRevE.68.026702} within PGMC. The idea is to suppress or prevent the worm from backtracking itself, letting fewer elementary actions be wasted on reversing what has been done. In Refs.~\cite{PhysRevE.66.046701,PhysRevE.68.026702} this is achieved by carefully solving a set of coupled equations. Here, we simply remove the actions associated with the $m$ sites last visited by the worm head (including the current one) from the elementary action space. For instance, if $m = 2$, both the spin at the current position of the head, $i_t$, and the spin at the previously visited site (if it exists), $i_{t-1}$, are excluded from possibly being flipped back in the next step: $\mathcal{A}_{\v \sigma, i_t} \gets \mathcal{A}_{\v \sigma, i_t} \setminus \bigl\{a_{\sigma_{i_t}}\inv, a_{\sigma_{i_{t-1}}}\inv\bigr\}$. 

It should be stressed that including a time dependence or \textit{memory} restricted to within an action is perfectly valid in an MCMC simulation; the elementary policies do \emph{not} need to be Markovian, as long as the entire chain policy \emph{is}~\footnote{This is not unique to the stochastic chain policies presented in this work. All traditional cluster algorithms are basically based on this fact. After all, they may themselves be cast as stochastic chain policies within the PGMC framework.}.

We test two worm policies with memory~\footnote{These are the only two worm policies with memory, apart from $m=1$, that makes sense in the kagome lattice simulations. Longer memories either do not add extra benefits, or run the risk of trapping the worm head in a position it cannot escape.},
\begin{itemize}
 \item[$\pi_{10}.$] As $\pi_9$, but with memory $m=2$, hence with immediate backtracking blocked.
 \item[$\pi_{11}.$] As $\pi_9$, but with memory $m=3$. Now, no more than 3 consecutive spin flips may happen within an elementary triangle of the kagome lattice.
\end{itemize}
(By construction, $\pi_9$ has $m=1$.) $\pi_{10}$ and $\pi_{11}$ do not result in any significant change from the already excellent sampling in the kagome ice regime by $\pi_9$, but the performances in the other parameter regimes are improved. See \cref{fig:MCMC_performance}. 

Now, in light of the criticism of hand crafted effective models within EMMC [\cref{sec:introduction}], it is not unreasonable to be equally reluctant towards hand crafted \emph{policies}, like the worm policies just presented. Sure, the physics of \cref{eq:Ising_model} is well known, making targeted policy modeling possible. But what if it was \emph{not}? The final policy model to be presented attempts at tackling this issue.  

With the success of the worm policies in mind, the idea is rather straightforward: Remove the special worm constraint, but keep the much more general memory constraint. In other words, let the policy be a stochastic chain policy with an elementary action space,
\begin{equation}
 \mathcal{A}_{s,t} = \mathcal{A}_s \setminus \left\{a_{t-\delta}\inv \mid \delta\in\{1,\ldots,m\}\right\}, \quad a_{t \le 0} = \varnothing
 \label{eq:STSA_action_space}
\end{equation}
where $\mathcal{A}_s$ is the space of all possible elementary actions, and $a_1, a_2, \ldots$ are the elementary actions taken so far. The constraint on immediately reversing elementary actions forces the policy to explore state space, instead of falling into the computationally wasteful local minima (attractive fix points) of bouncing. We call such a policy a \textit{short term self avoiding} policy with memory $m$. 

To demonstrate the short term self avoiding idea, we simply replace the $\pi_\text{move}$ elementary policy of the worm algorithms $\pi_9$--$\pi_{11}$ with a new $\pi_\text{SA}$ self avoiding elementary policy. The spin flip preference of $\pi_\text{SA}$ is given by $h_4$, \cref{eq:h_4}, and instead of using just a constant preference for terminating the update, we choose to use a slightly more involved -- and flexible -- stopping preference,
\begin{equation}
  h_\text{stop}(a_\varnothing|\v \sigma) = \ln\left[\sum_{i = 1}^N \exp(\theta^\text{stop}_{q_{i}}) \right],
  \label{eq:h_stop}
\end{equation}
parametrized by $\v \theta^\text{stop}$. As for $h_4$, the local feature map $q_i$ is also given by \cref{eq:feature_map}. In total, there are $29$ independent parameters. 

Four such short term self avoiding policies are tested:
\begin{itemize}
 \item[$\pi_{12}$.] With memory $m=1$.
 \item[$\pi_{13}$.] With memory $m=3$.
 \item[$\pi_{14}$.] With memory $m=6$.
 \item[$\pi_{15}$.] With memory $m=10$.
\end{itemize}
As seen in \cref{fig:MCMC_performance}, avoiding backtracking actions is indeed the crucial ingredient in improving the Markov chain dynamics, with $\pi_{12}$--$\pi_{15}$ performing better than any of the other non-specialized policies in the deep spin ice regime. The short term self avoiding policies may not be quite as efficient as the worm policies here, but, crucially, their aptness come from machine learning, rather than a hard coded design. 

\begin{figure}
 \includegraphics[width=\columnwidth]{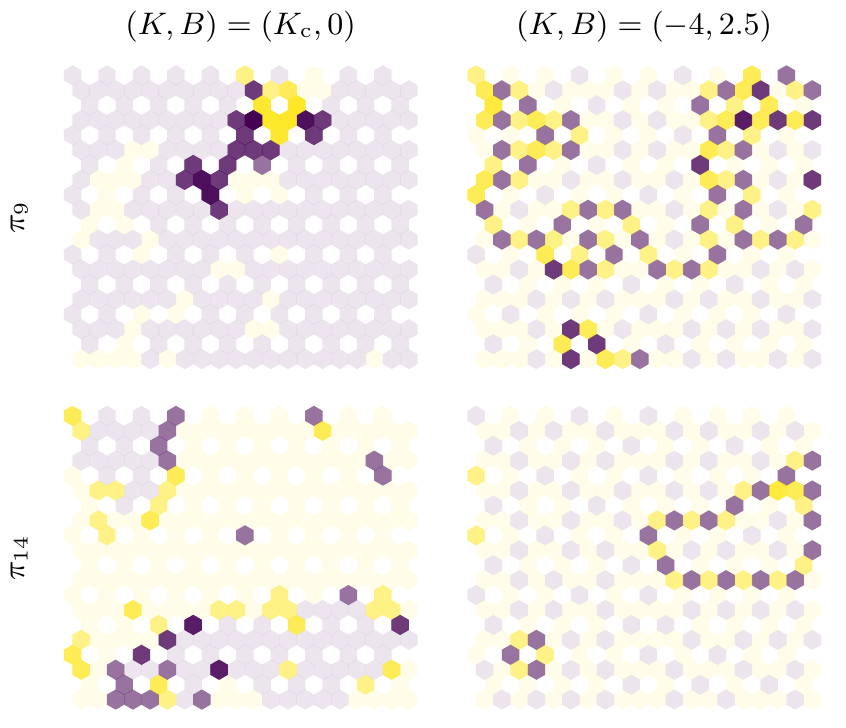}
 \caption{Snapshots of configurations and most recent actions leading to them, at the critical point (left) and in the kagome spin ice regime (right). In the upper panels, the actions have been selected by the worm policy $\pi_9$, while in the lower panels, they have been selected by the short term self avoiding policy $\pi_{14}$. Stronger intensity indicate that the same spin has been flipped (and flipped back) multiple times within the last action, while the weakly colored background spins have not been flipped at all. Note that while the elementary actions of the worm policy, by construction, form a single connected set, the actions of the short term self avoiding policy have no such restrictions imposed. This makes the latter more flexible in adjusting to other parameter regimes, where string or loop updates may no longer be optimal.
 }
 \label{fig:snapshot_of_updates}
\end{figure}

\Cref{fig:snapshot_of_updates} compares updates generated by the worm policy $\pi_{9}$ and the short term self avoiding policy $\pi_{14}$ in the critical and frustrated regimes. In the latter, both construct efficient, non-local loop updates. In the critical regime, however, the worm policy is no longer particularly advantageous; the stringy worm updates do not cope well with the fractal nature of the critical configurations, resulting in inefficient updates containing multiple bounce processes. The short term self avoiding policy, on the other hand, does not suffer from the restricted action space of the worm policy, and may therefore to a greater degree adjust to the experienced physics. 

It is not unreasonable to speculate that a policy model even more flexible than $\pi_{14}$ could ``rediscover'' a cluster algorithm~\cite{PhysRevLett.58.86,PhysRevLett.62.361} at criticality, as the cluster algorithms may themselves be regarded as stochastic chain policies with memory. If so, a single policy model would be sufficient to competitively simulate \cref{eq:Ising_model} in the entire $(K,B)$ space.

The last to be examined in this work, the short term self avoiding policy model certainly do not represent the end of the story. The need for selecting the ``right'' memory -- too short and the benefits of self avoidance are not maximized, too long and the policy is no longer able to adequately track the desired probability distribution -- again means tedious and inefficient hyperparameter tuning. It is not hard to envision better and more flexible policies that deals with such issues -- and more; after all, the stochastic chain policies with memory presented constitute only a few members of a broader class of policies with history dependent elementary policies. In principle, such a history dependence can also be modeled by a differentiable structure and subsequently machine learned. The practicalities, however, are left for future work.

\subsection{PGMC for other systems}\label{sec:other_systems}

While ideas like the chain and self avoiding policies do not depend on the specifics of the probability weight model in question, other aspects of the preceding examples do. Specifically, the (elementary) policy models of the kagome Ising model do all exploit the translational invariance of \cref{eq:Ising_model} and the binary nature of the Ising spins -- both nonuniversal, case dependent traits. Even though the PGMC idea is generally applicable, specific policies may not be.

Clearly, if symmetries like translational invariance are not present in the system of interest, this should be reflected in the policy model. Not only is this likely to render the training stage more expensive (there is less ``information reuse'' when symmetries are not imposed), a feature map approach, as the one employed in this work, may no longer be computationally efficient or feasible. Then, other action selection strategies may be necessary, like those briefly mentioned in \cref{sec:modeling_the_policy} [see also \cref{app:simulation_details}].

Furthermore, if the degrees of freedom are not binary, it is no longer sufficient to let the elementary actions be of the form of ``flips'', i.e. ``change the value of some degree of freedom $i$ to its other value''. In cases where a degree of freedom still constitute a finite, discrete set $\chi$ (e.g. $\chi = \mathbb{Z}_n$ with $n > 2$) it may be possible to construct (chain) actions on the form ``select degree of freedom $i$ and change it from $\sigma_i$ to $\sigma_i'$, $\sigma_i,\sigma_i' \in \chi_i$''. Assigning individual weights to all elements of the (local) action space is, however, only viable up to some point. Beyond this, for cases where the number of elements in $\chi$ is too large or even infinite (i.e. for continuous degrees of freedom, like $\chi = \mathbb{R}$, $\chi = \mathrm{O}(2)$, etc.) some degree of probability weight/density ``interpolation'' has to take place. 

As an example, in the case of a classical XY model ($\chi = \mathrm{O}(2)$), each degree of freedom can be parametrized by a continuous angle $\phi \in [0,2\pi)$. A possible policy for updating a single degree of freedom may then read
\begin{enumerate}
 \item Select a site $i$ with probability $\pi(i|s)$.
 \item Then, select a sector $\chi_k \equiv [\frac{2\pi k}{n}, \frac{2\pi (k+1)}{n})$, $k \in \{0,1,\ldots,n-1\}$, with probability $\pi(\chi_k|\text{neighborhood of $i$})$.
 \item Finally, select a new angle $\phi_i' \in \chi_k$ by sampling a uniform random number $R \in [0,1)$ and then calculate $\phi_i' = \frac{2\pi(k+R)}{n}$.
\end{enumerate}
The two first steps can then be modeled (parametrized by $\v \theta$) and learned in a way similar to what was done for the kagome Ising model of the previous subsections~\footnote{There are, obviously, infinitely many alternative -- and probably better -- policy models in this particular case. Determining a good one is outside the scope of this work.}. Since the infinitesimal ``probability weight'' of step 3 is constant and independent of state, its value is not of importance when calculating \cref{eq:gradient_log_pf}.

Interestingly, systems of continuous degrees of freedom may possess additional ``geometric'' information not available in the discrete systems. Thus, for systems of continuous degrees of freedom, one may contemplate policies where not only the state, but also properties like the state space gradient of the log probability (or even a machine learned effective model) is taken into account. Such a scheme could then be considered a PGMC ``equivalent'' to the Metropolis-adjusted Langevin algorithms (MALA)~\cite{RobertsTweedie:1996,GirolamiCalderhead:2011,Durmus:2017}.

\section{Discussion}\label{sec:discussion}

After having seen some of PGMC's potential in the case studies of the previous section, we now turn back to the more general perspective. The aim of the following brief discussions is to put PGMC somewhat in context with respect to MCMC in general and EMMC in particular. We also mention a few challenges and possible pitfalls concerning PGMC. 

\subsection{Machine learning MCMC}

Common to all traditional Monte Carlo algorithms is the trade-of between the inclusion of a priori knowledge and stochasticity, specialization and generality: A very general -- hence also ``oblivious'' -- Monte Carlo algorithm, which is applicable to many problems, will typically perform poorer than a specialized, but limited algorithm that exploits many aspects of the problem at hand. Intriguingly, MCMC methods incorporating machine learning, like EMMC and PGMC, have the potential to bridge this gap: By letting the algorithms automatically learn from and adjust to the sampled distribution, these methods may, in a sense, be both general and specialized at the same time; general in usability, specialized in computation. 

The value of such a property should not be understated. Even if algorithms hard coded by domain experts may surpass the performance of machine learned algorithms in some cases, the usefulness of having general, flexible, and \emph{still} reasonably performant tools is significant in many real world scenarios. 

Looking further, there is nothing intrinsic that prevents the machine learning methods from surpassing what is already known. In fact, for anything beyond the simplest and most well understood models, machine learning methods are likely to be a practical necessity if one wish to reap the benefits of specialization.

\subsection{EMMC and PGMC}

The main difference between the EMMC and PGMC methods lies in their computational perspectives: While EMMC is \textit{static}, exchanging one model (the original one) for another (the effective), PGMC is \textit{dynamic}, attempting to model the Markov chain dynamics directly. Where the effective models of EMMC are constructed by imitating the training data, the policy search of PGMC aims at surpassing what is currently known. The autocorrelation -- a central trait of the final MCMC sampling -- is specifically taken into account in the PGMC policy optimization, while it is disregarded in EMMC's effective model training. 

The PGMC scheme offers other advantages in the optimization stage, both conceptually and practically: The goal is clearly and quantitatively stated (``maximize the expected performance factor''), regardless of implementation details. This provides an explicit path forward, leading directly to efficient real-world algorithms -- as seen in \cref{sec:PGMC_in_practice}. Efficient bootstrapping techniques, like on-policy learning, is a natural consequence of the dynamical perspective.

The \textit{global view} that readily comes with the policies is another asset of PGMC: computational resources can easily be directed towards the most important or relevant degrees of freedom of a state. Exemplified in \cref{fig:single_flip_PGMC_snapshots}, even a relatively simple policy may target structures like domain walls or particle like excitations, while spending less time on fluctuations of minor importance. Thus, a PGMC simulation can easily approach a performance that would otherwise require more specialized and involved rejection free MCMC schemes.

The situation is less clear for EMMC. Even with an effective model in place, there is still a need to find or design good updates, with no definite guidelines to follow. The effective model updates used are therefore prone to be either generic and relatively poor, or specialized and of limited applicability.

\subsection{Pitfalls and challenges}\label{sec:challenges}
In machine learned MCMC, as in reinforcement learning, there is danger in trading stochastically sampled training data for increased simulation speed: If the available information is inadequate or wrong, the effective models or policies -- hence the final sampling -- may be bad. In PGMC, extra care has to be taken when the policy training is conducted online, on-policy. The reason is that the optimization may, in some situations, enter a feedback loop where the policy, with an increasingly strong bias, selects update proposals among only a subset of relevant states. The exploration of the state space stops and, for all practical purposes, ergodicity is broken. Taking the worm PGMC of \cref{sec:stochastic_chain_policies} as an example: if the policy training was not properly controlled, it could happen that the learned policy would end up staying entirely within the kagome spin ice manifold. The correct sampling at $(K,B) = (-4,2.5)$ should contain a low density of excited states. Nevertheless, these excitations may not always have been encountered sufficiently often during a crucial period of the training, leading the policy into a ``spiral of suboptimization'' where ``excited states where not proposed because excited states were not proposed''.

Such problems, although mostly absent from the simple test cases of \cref{sec:PGMC_in_practice}, are likely to happen more if policy model complexity, hence also (overfitting) \textit{capacity}, is increased. A possible solution, similar to what is done within reinforcement learning~\cite{SuttonBarto2018}, could be to move away from strict on-policy based learning by including some stochastic noise in the behavior policy. For example, a uniformly random action could be selected with some low frequency, sacrificing a little bootstrapping performance for a persistent exploration of the state space.

Another set of issues may arise with the training of intricate policies, e.g. chain policies with long actions. First, a larger policy model requires more information and is typically more demanding to train. Second, the longer actions take longer to construct, hence the frequency of generating training samples is reduced. Furthermore, the longer delay between selecting elementary actions and evaluating the total action means that the propagation of information back through the chain will be less efficient and more noisy, as the effect of later actions may strongly depend on earlier ones. Overall, the positive aspects of having more complex, globally updating policies are also the ones that makes it harder to train them. As such, the issues cannot easily be circumvented. Fortunately, however, many of these sides of policy training are well known and tackled within the reinforcement learning community, with solutions ripe for being incorporated into PGMC: It could, for instance, be fruitful to use information acquired \emph{during} the updates to build \textit{value} or \textit{action-value} function models, either to support the policy model training, or as an alternative to the policy models discussed in this work~\cite{SuttonBarto2018}.

Finally, the stability and convergence speed of the training stage can likely be improved by storing and reusing more of previously obtained data in an offline, off-policy approach (\textit{experience replay}), like demonstrated in Ref.~\cite{Mnih2015}. With a slight modification, such a method could also benefit from simulation data obtained with slightly different models (e.g. the model of interest, but at a different temperature), analogous to what is achieved with reweighting techniques~\cite{PhysRevLett.61.2635,PhysRevLett.63.1195}.

\section{Conclusion}\label{sec:conclusion}
In this paper, we have seen how reinforcement learning and Markov chain Monte Carlo simulations may be combined in a general, unbiased, and powerful framework: Policy Guided Monte Carlo. The basic theory of PGMC has been developed and subsequently scrutinized, with the aim of advancing PGMC as a practical tool for both improving and expanding the scope of MCMC sampling, as well as automatizing the process of developing new, efficient MCMC algorithms. Examples of increasing sophistication, centered around a simple, but challenging Ising model on the kagome lattice, underline the effort and show how PGMC may be done in practice. We also discuss PGMC's relation to other MCMC algorithms and where care has to be taken.

The gains of PGMC may span the entire spectrum, from modest speedups in handling known problems to novel samplers tackling new ones. We therefore expect PGMC to be widely beneficial in the realm of MCMC simulations.

\begin{acknowledgments}
The author would like to thank Shigeki Onoda, Yukitoshi Motome, and Yasuyuki Kato for fruitful discussions and feedback related to this work. He would also like to thank an anonymous referee for bringing the adaptive MCMC and MALA algorithms to his attention.

This work was supported by Grants-in-Aid for Scientific Research (KAKENHI) (number JP16H02206).
\end{acknowledgments}

\appendix
\section{Simulation details}\label{app:simulation_details}

\subsection{Implementation}

All implementations were done in the \textsc{Julia}~\cite{BezansonETAL:2017} programming language with the help of the \textsc{Flux} machine learning library~\cite{Innes:2018} and its implementation of the ADAM stochastic gradient decent algorithm~\cite{KingmaBa:2014}. 

\begin{figure}
  \includegraphics[width=0.8\columnwidth]{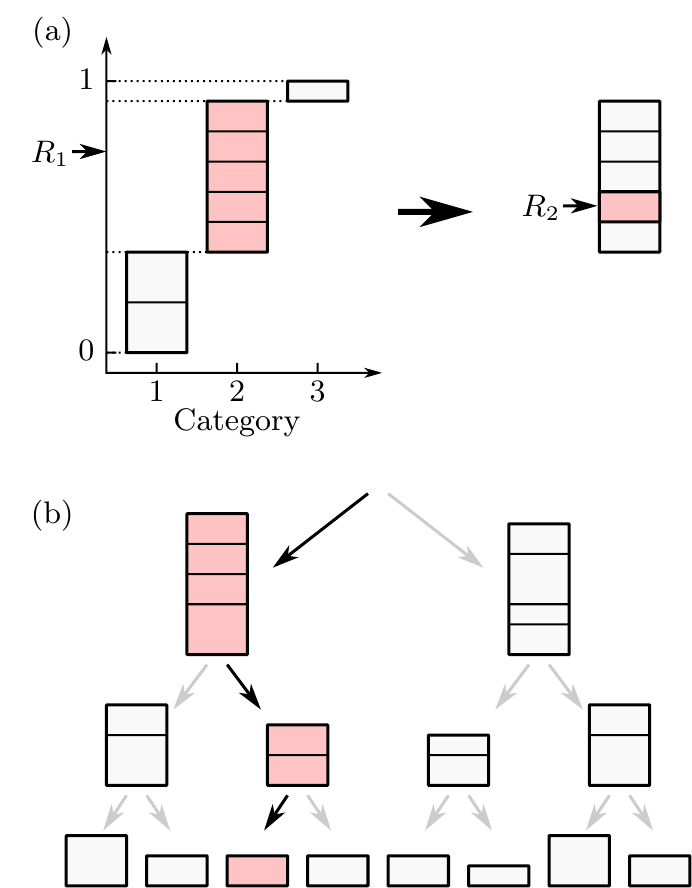}
 \caption{(a) Inverse transform and (b) weighted binary tree selection of an action based on some policy $\pi$. Each box represents some action $a$, with the height being given by $\pi(a)$. In (a), the action space is sorted into feature categories. A category is chosen (colored) according to the total marginal probability of the category by drawing a uniform random number $R_1$. Next, a specific action is selected uniformly at random among the members of the category by random number $R_2$. The total number of operations is constant with respect to the size of the action space, $N_{\mathcal{A}}$. In (b), the probability (weights) are arranged in a binary tree structure, with the weight of each parent node being the sum of its children nodes. An action is selected by traversing the tree from root to leaf, each time selecting the left or the right child at random, proportionally to their relative weights. This scheme comes at the (mild) expense of an $\order{\log N_{\mathcal{A}}}$ scaling in the number of operations needed. The advantage is that it can be efficiently used even when the number of possible categories is large or even infinite. 
 Naturally, when an action has been applied to a state, both the category groupings in (a) and the weights of the tree in (b) will have to be updated. In cases with ``short range feature maps'' (i.e. when the features only depend on a finite neighborhood of the degree of freedom being updated by the action) this can also be done with $\order{1}$ and $\order{\log N_{\mathcal{A}}}$ number of operations, respectively.
 }
 \label{fig:action_selection}
\end{figure}

Actions were efficiently chosen by first picking a feature category using inverse transform sampling and then selecting among the equally weighted members of the category. An alternative approach of using a weighted binary tree was also successfully tested, although not used in the test simulations presented here. Both methods are illustrated in \cref{fig:action_selection}.

Naturally, the complete information about the actions had to be temporarily stored, both for calculating the policy value of the inverse action and for being able reconstruct the previous state in case of rejection.

\subsection{Simulation parameters used}
In all the displayed simulation results, the system size was set to $L=10$ and the spins were initialized at random. The policy parameters were set to $\v \theta = \v 0$ before training. The other simulation parameters were chosen as to obtain reliable results, e.g. clear convergence and more-than-sufficient equilibration. \Cref{fig:MCMC_performance,fig:spins_flipped_frequency} show averages and standard deviations based on 16 completely independent simulation runs. The simulation parameters of \cref{fig:MCMC_performance} are listed in \cref{tab:simulation_parameters}. In \cref{fig:spins_flipped_frequency}, a simulation run consisted of $200N$ training iterations, $1000N$ equilibration iterations, and $10^7$ samples (without thinning).
\begin{table}
  \caption{\label{tab:simulation_parameters} Simulation parameters used in each of the 16 independent simulation series behind each datapoint presented in \cref{fig:MCMC_performance}. \textit{Equilibr.} stands for \textit{equilibration}. A thinning factor of $x$ means that only one out of every $x$ MCMC iteration steps is sampled. (The time scale is still set by a single iteration.)}
  \begin{ruledtabular}
  \begin{tabular}{ccrrrr}
  $(K,B)$		& Policies 	& \specialcell{Training\\iterations} & \specialcell{Equilibr.\\iterations} & \specialcell{Thinning\\factor} & Samples\\
  \colrule
  $(K_\text{c},0)$ 	& $\pi_1$ 		& --- 		& $5000N$ 	& $50N$ 	& $5000$ 	\\
			& $\pi_2$--$\pi_4$ 	& $100N$ 	& $100N$ 	& $50N$ 	& $5000$ 	\\
			& $\pi_5$--$\pi_7$ 	& $100N$ 	& $100N$ 	& $10N$ 	& $10000$ 	\\
			& $\pi_8$ 		& $200N$ 	& $100N$ 	& $10N$ 	& $10000$ 	\\
			& $\pi_9$--$\pi_{11}$ 	& $500N$ 	& $1000$ 	& $100$ 	& $5000$ 	\\
			& $\pi_{12}$--$\pi_{15}$& $100N$ 	& $1000$ 	& $100$ 	& $5000$ 	\\
  $(0.5,1)$ 		& $\pi_1$ 		& --- 		& $500N$ 	& $1N$ 		& $5000$	\\
			& $\pi_2$--$\pi_4$ 	& $100N$ 	& $100N$ 	& $15$ 		& $5000$ 	\\
			& $\pi_5$--$\pi_7$ 	& $100N$ 	& $100N$ 	& $1$ 		& $50000$ 	\\
			& $\pi_8$ 		& $200N$ 	& $100N$ 	& $1$ 		& $50000$ 	\\
			& $\pi_9$--$\pi_{11}$ 	& $100N$ 	& $1000$ 	& $1$ 		& $10000$ 	\\
			& $\pi_{12}$--$\pi_{15}$& $100N$ 	& $1000$ 	& $1$ 		& $10000$ 	\\			
  $(-4,0)$ 		& $\pi_1$ 		& --- 		& $100N$ 	& $1N$ 		& $2000$	\\
			& $\pi_2$--$\pi_4$ 	& $100N$ 	& $100N$ 	& $100$ 	& $5000$ 	\\
			& $\pi_5$--$\pi_7$ 	& $100N$ 	& $100N$ 	& $20$ 		& $5000$ 	\\
			& $\pi_8$ 		& $200N$ 	& $100N$ 	& $20$ 		& $5000$ 	\\
			& $\pi_9$--$\pi_{11}$ 	& $200N$ 	& $1000$ 	& $5$ 		& $10000$ 	\\
			& $\pi_{12}$--$\pi_{15}$& $200N$ 	& $1000$ 	& $1$ 		& $10000$ 	\\			
  $(-4,0.5)$ 		& $\pi_1$ 		& --- 		& $2000N$ 	& $50N$ 	& $2000$ 	\\
			& $\pi_2$--$\pi_4$ 	& $100N$ 	& $1000N$ 	& $2N$ 		& $10000$ 	\\
			& $\pi_5$--$\pi_7$ 	& $100N$ 	& $1000N$ 	& $N$ 		& $5000$ 	\\
			& $\pi_8$ 		& $200N$ 	& $1000N$ 	& $N$ 		& $5000$ 	\\
			& $\pi_9$--$\pi_{11}$ 	& $300N$ 	& $1000$ 	& $5$ 		& $10000$ 	\\
			& $\pi_{12}$--$\pi_{15}$& $200N$ 	& $1000$ 	& $5$ 		& $10000$ 	\\			
  $(-4,2.5)$ 		& $\pi_1$ 		& --- 		& $5000N$ 	& $50N$ 	& $2000$ 	\\
			& $\pi_2$--$\pi_4$ 	& $100N$ 	& $2000N$ 	& $500N$ 	& $1000$ 	\\
			& $\pi_5$--$\pi_7$ 	& $100N$ 	& $2000N$ 	& $10N$ 	& $5000$ 	\\
			& $\pi_8$ 		& $200N$ 	& $2000N$ 	& $10N$ 	& $5000$ 	\\
			& $\pi_9$--$\pi_{11}$ 	& $300N$ 	& $1000$ 	& $1$ 		& $10000$ 	\\
			& $\pi_{12}$--$\pi_{15}$& $200N$ 	& $1000$ 	& $1$ 		& $10000$ 	\\			
  \end{tabular}
  \end{ruledtabular}
\end{table}

\bibliography{references.bib}

\end{document}